\documentclass[pra,twocolumn,superscriptaddress,showpacs,amsmath,amstex,amssymb,citeautoscript]{revtex4-1}

\usepackage[T1]{fontenc}
\usepackage[utf8]{inputenc}
\usepackage{lipsum}
\usepackage{amsmath}
\usepackage{amssymb}
\usepackage{bm}
\usepackage{bbm}
\usepackage{braket}
\usepackage{xcolor}
\usepackage{pifont}
\usepackage[mathscr]{euscript}
\usepackage[shortlabels]{enumitem}
\usepackage{tabularx}
\usepackage{graphicx}
\usepackage{ragged2e}

\usepackage{graphicx}
\usepackage{lipsum}
\usepackage{amsmath, bm}
\allowdisplaybreaks
\usepackage{float}

\usepackage{dsfont}
\usepackage{comment}

\usepackage[colorlinks=true]{hyperref}  
\hypersetup{
    bookmarks=true,         
    unicode=false,          
    pdftoolbar=true,        
    pdfmenubar=true,        
    pdffitwindow=false,     
    pdfstartview={FitH},    
    pdftitle={High-harmonic generation from chiral Bloch states},    
    pdfauthor={Almeida, Kort-Kamp, Scheurer},     
    pdfsubject={},   
    pdfcreator={},   
    pdfproducer={}, 
    pdfkeywords={} {} {}, 
    pdfnewwindow=true,      
    colorlinks=true,       
    linkcolor=blue, 
    citecolor=blue,        
    filecolor=magenta,      
    urlcolor=blue           
}

\newcommand\numberthis{\addtocounter{equation}{1}\tag{\theequation}}
\newcommand{\equref}[1]{Eq.~(\ref{#1})}

\newcommand{\secref}[1]{Sec.~\ref{#1}}
\newcommand{\figref}[1]{Fig.~\ref{#1}}
\newcommand{\refcite}[1]{Ref.~\onlinecite{#1}}

\newcommand{\appref}[1]{Appendix~\ref{#1}}

\newcommand{\diff}{\mathrm{d}}

\renewcommand{\vec}[1]{\boldsymbol{\mathrm{#1}}}

\definecolor{wrongultramarine}{rgb}{1,0.5,0}
\begin{document}

\title{High-harmonic generation in systems with chiral Bloch states: \\ application to rhombohedral graphene}

\author{Jessica O. de Almeida}
\email{jessica.almeida@itp3.uni-stuttgart.de}
\affiliation{Institute for Theoretical Physics III, University of Stuttgart, 70550 Stuttgart, Germany}

\author{Wilton J. M. Kort-Kamp}
\affiliation{Theoretical Division, Los Alamos National Laboratory, Los Alamos, NM 87545, United States}

\author{Mathias S.~Scheurer}
\affiliation{Institute for Theoretical Physics III, University of Stuttgart, 70550 Stuttgart, Germany}

\begin{abstract}
Nonlinear light-matter interaction and, in particular, 
high-harmonic generation (HHG) are fundamentally interesting and frequently discussed as versatile probes of quantum materials with potential for optical information processing applications. Meanwhile, there has also been significant progress in graphene-based multilayer systems to engineer interesting band structures and boost correlation effects. Motivated by the successful demonstration of HHG in graphene, we here study this effect in rhombohedral stacks of $n$ layers of graphene, a recent very prominent representative of correlated multilayer graphene systems. We show how the chiral Bloch states of the valleys of this system crucially affect the HHG. The ``winding’’ of the Bloch states scales linearly with $n$, just like the dominant harmonic order. The location of the strongest quantum geometry in momentum space on a ring of finite radius is shown to be imprinted on the time-dependent momentum distribution at the beginning of the strong laser pulse. We further demonstrate that the presence of an interaction-induced splitting of the two valleys leads to a complex interplay of the opposite chiralities of the two valleys, directly visible in the $n$ dependence of the circular dichroism. We also analyze the impact of doping and identify a quantity that tracks the net chirality of the occupied states. Our findings show that rhombohedral graphene constitutes a promising platform for exploring rich nonlinear optical phenomena. 
\end{abstract}

\maketitle

\section{Introduction}
In recent years, the systematic study of effects rooted in the non-trivial momentum dependencies of multi-component Bloch states---often collectively referred to as quantum geometry---has become a very active field of study in condensed matter physics, relevant to a broad variety of phenomena \cite{QGReview,Jiang_2025,QGSuperfluid,QGScales}. This is particularly true in systems with low symmetries and, hence, fewer constraints on the associated matrix elements. For instance, if all mirror symmetries and time-reversal symmetry are broken, the Bloch states are chiral and exhibit a non-zero net Berry curvature $\Omega$. In this regard, rhombohedral multi-layer graphene (RMG), where graphene layers are stacked such that the B sublattice of one layer is just below the A sublattice of the subsequent layer, see \figref{fig:intro}(a), provides a particularly exciting and recent example. Without any two-fold rotational symmetry $C_{2z}$ or inversion (which combined with time-reversal would lead to an anti-unitary intravalley symmetry, prohibiting any Berry curvature) in the crystal structure, the Bloch states of RMG are chiral with $\Omega \neq 0$. Interestingly, $\Omega(\vec{k})$ and quantum geometric effects in general are concentrated on a ring of finite radius in momentum ($\vec{k}$) space encircling the $\textrm{K}$ and $\textrm{K}’$ points and the magnitude is tunable by the number $n$ of rhombohedrally stacked layers. In combination with the fact that this system exhibits strongly correlated physics \cite{SCTrilayer,HalfQuarterMetals,NontwistedReview,valleySCExp,JiaSixLayers}, including the spontaneous emergence of an imbalance between the two valleys, which breaks time-reversal symmetry, RMG has attracted a lot of interest \cite{AcousticPhononsDasSarma,BergRGKohnLuttinger,YiZhuangKohnLuttinger,BitanRoy,ShubhayusPaper,MultilayerGeneralSerbyn,PacoBiAndTri,MacDonaldFRG,Paco2023,Chubukov,DongSpinCanting,AliceaQuantumGeometry,StandfordWithTheRelativeChirality,Paco,YahuisPaper,OurEnergetics,2025arXiv251019943S,FranzSCRingOfFire,PatrickLeeControl,BerryTrashcan}, including in the context of quantum geometry. 

Harmonic generation, a non-linear optical phenomenon where the driving of a system by a high-intensity laser at a fundamental frequency $\omega_0$ leads to the emission at frequencies $\omega(l) = l \omega_0$, with integer $l>1$, has also attracted significant attention in recent years \cite{PhysRevA.49.2117,corkum1993plasma,lewenstein1994theory,yue2022introduction,vampa2017merge,Review2DHG,AnotherReview2D,zhou2022engineering}. Apart from its technological relevance, one of the central reasons is its relation to the aforementioned quantum geometry and topology of Bloch bands \cite{PhysRevLett.129.227401,PhysRevB.102.081121,kim2022theory,malla2023ultrafast,chacon2020circular,heras2026pulse,luu2018measurement,silva2019topological,Wu2015High,bai2021high,bauer2018high,heide2022probing,bera2023topological,ciappina2025solid}. The relevance of Bloch-state matrix elements arises because the light-matter interaction involves the band projection of the dipole operator. Harmonic generation, including with large $l$ (HHG), is also very actively studied in two-dimensional systems \cite{Review2DHG,AnotherReview2D,zhou2022engineering}, such as graphene---both experimentally \cite{hafez2018extremely,yoshikawa2017high,soavi2018broadband,cha2022gate,baudisch2018ultrafast,molinero2024high,ExperimentSHG,yoshikawa2019interband,SHG_differentstackings} and theoretically \cite{PhysRevB.110.054103,mrudul2021high,sato2021high,chen2019circularly,rakhmanov2025optical,boyero2022non,PlasmonAssisted,D0TC02036B,murakami2022doping,ikeda2020high,avetissian2022efficient,guan2023optimal,dong2021ellipticity, zhang2021orientation, zhang2021orientation,rana2022generation}.

This motivates us to study the physics of harmonic generation in RMG for different numbers $n$ of graphene layers. We elucidate the rich physics coming from the low symmetries of the system, in particular the lack of $C_{2z}$ symmetry, a crucial difference compared to the related flat-band system twisted bilayer graphene \cite{ikeda2020high}, and demonstrate the key role of the quantum geometry of the Bloch states for harmonic generation. We investigate the consequences of an interaction-induced valley imbalance and doping for the HHG. As a sensitive measure of the valley-resolved physics, we employ circular dichroism (CD) of our harmonic generation spectrum, which is widely used as a probe of chirality (see, e.g., \cite{asteria2019measuring,ranjbar2009circular}) and has also been proposed as a way of accessing Chern numbers \cite{chacon2020circular,kim2022theory,heras2026pulse,silva2019topological}. We find that the competition of the two valleys with different interaction-induced gaps leads to a sign change of the dominant circular dichroism signature as a function of the number $n$ of graphene layers.

The remainder of the paper is organized as follows. In \secref{ModelAndApproach}, we introduce the basic model and formalism that we use. To illustrate the key physics in a minimal setting, we first start our analysis of HHG spectra only taking into account one of the two valleys of RMG, see \secref{SingleValleyLimit}. In \secref{IncludingSecondValley}, we incorporate both valleys and their interplay. The results are summarized in \secref{Conclusion}. Multiple appendices provide additional data and analytical details.

\section{Model and Approach}\label{ModelAndApproach}
We begin this section by introducing the effective two-band model that is used to describe the low-energy physics of RMG, which can be more generally thought of as a minimal prototype of a chiral continuum model. We will couple it to an external electromagnetic field associated with the applied strong pulsed laser and outline how we compute the resultant HHG.

\subsection{Bloch Hamiltonian of RMG}
\begin{figure}[tb]
    \centering
    \includegraphics[width=\columnwidth]{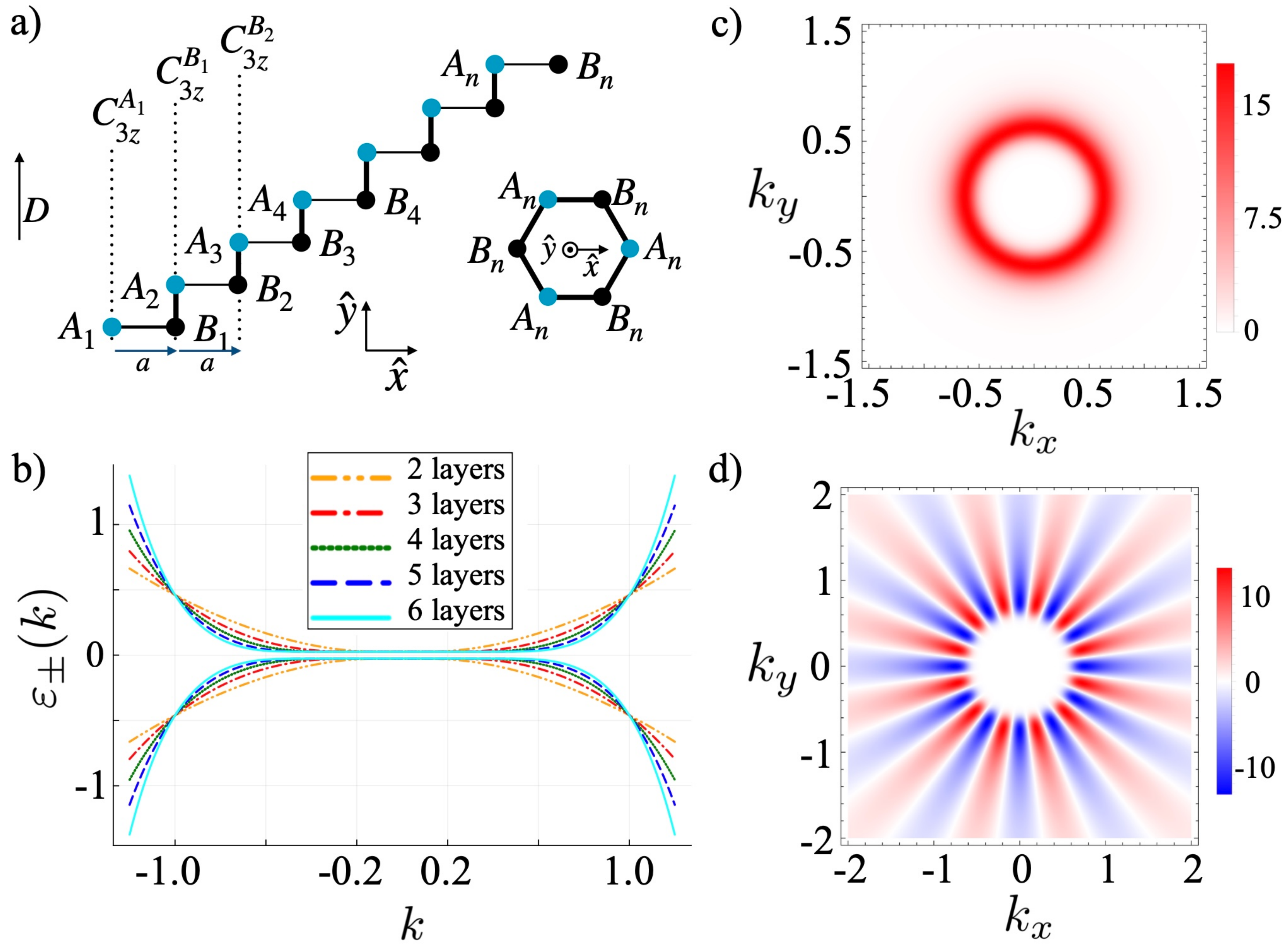}
    \caption{\justifying a) Side-view of rhombohedral stack of $n$ graphene layers, i.e., sublattice $A_{j+1}$ of the $(j+1)$th layer is superposed with site $B_j$ of the $j$th layer; $D$ is the applied displacement field, which creates a potential off-set between the layers and the inset on the right shows one honeycomb of one graphene layer. b) Electronic dispersion, \equref{eq:Energy}, of the effective two-band model of RMG we use, for $n=2-6$. c)~Berry curvature $\Omega^{(\xi=+)}_{+}(\vec{k})$ for $n=6$. d) Square of the off-diagonal components of the dipole matrix in \equref{eq:DmomEl}, $[\vec{d}^{(\xi=+)}_{+-}(\bm{\mathrm{k}})]^2$, with $n=6$.
    It shows a flower pattern, with $2n$ ``petals'' (blue/red). Parameters for (b)-(d) and in our numerics are  $w=0.026\,$a.u., $\gamma_1=0.46\,$a.u.. Momenta are measured in units of $k_c$, such that effectively $k_c=1\,$a.u..}
    \label{fig:intro}
\end{figure}

We here consider rhombohedral stacks of $n$ graphene layers, i.e., the honeycomb lattices with the two sublattices denoted by $A_{j}$ and $B_{j}$ in the $j$th graphene layer, are stacked such that $B_{j}$ and $A_{j+1}$ are vertically on top of each other, see \figref{fig:intro}(a). An important experimental tuning parameter of the system is an external displacement field, denoted by $D$, which effectively creates a potential offset between neighboring layers. Using a full tight-binding model on the honeycomb lattice would give rise to a total of $2n$ bands (per spin). However, near charge neutrality, there are only two low-energy bands---one just above and one below the Fermi level in the vicinity of the $\textrm{K}$ ($\xi=+$) and $\textrm{K}'$ ($\xi = -$). The weight of their Bloch states is localized on the non-dimerized sites, i.e., the sublattices $A_1$ and $B_n$. This allows us to construct~\cite{koshino2009,koshino2010interlayer,PhysRevB.82.035409,min2008pseudospin,koshino2010parity,muten2021exchange} a minimal two-band model (per spin) for each of the two valleys $\xi=\pm$, with the Bloch Hamiltonian.
\begin{equation}
    h^{(n)}_{\xi}(\vec{k})=\left( \begin{matrix}
                        -\mu+w &  -\gamma_1 (-k^*_{\xi}/k_c)^{n}  \\
                       -\gamma_1 (-k_\xi /k_c)^{n} &  -\mu-w
                                    \end{matrix}	
                                \right).\\     
    \label{eq:Hamiltonian}
\end{equation} 
Here, $\vec{k}=(k_x,k_y)$ is the momentum deviation from the respective $\textrm{K}$ or $\textrm{K}'$ point and we introduced the complex momentum coordinate $k_\xi = k_x + i \xi k_y$. Besides, $\mu$ is the chemical potential, $w$ is controlled by the displacement field, $\gamma_1$ is the interlayer nearest neighbor hopping amplitude between superposed sites $B_{j}$-$A_{j+1}$, and the momentum scale is $k_c=2\gamma_1/(\sqrt{3}a\gamma_0)$. In the latter, $a$ is the intralayer distance between sublattices and $\gamma_0$ the intralayer nearest neighbor hoppings of the original tight-binding model.  

An important advantage of the effective two-band model in \equref{eq:Hamiltonian} is that the eigenstates and the associated matrix elements we will need below are readily derived analytically (see \appref{sec:AppExp}). The spectrum reads as
\begin{equation}
    \epsilon_{p}(\vec{k}) = -\mu + p \sqrt{w^2+\gamma_1^2 (k/k_c)^{2n}},
    \label{eq:Energy}
\end{equation}   
where $p=\pm$ labels the two bands and $k=\sqrt{k_x^2 + k_y^2} = |k_\xi|$. It is shown in \figref{fig:intro}(b) and reveals that, with increasing $n$, the bands become flatter in a range of momenta of order $k_c$. The associated increase in the density of states is also expected to be the reason why RMG displays multiple correlated phases, as revealed by recent experiments \cite{SCTrilayer,HalfQuarterMetals,NontwistedReview,valleySCExp,JiaSixLayers}.

Note that this minimal model leads to a dispersion that is even in momentum, $\epsilon_{p}(\vec{k}) = \epsilon_{p}(-\vec{k})$. This is also why the band energies are the same in both valleys $\xi=\pm$ and why we suppressed a valley index $\xi$ in \equref{eq:Energy}. While one could introduce additional trigonal warping terms to break the degeneracy, we will not do so here since this will allow us to focus on the effect of the Bloch wave functions $\ket{\Psi_{\vec{k},p}^{(\xi)}} = e^{i(\vec{k}-\xi\vec{K})\cdot \hat{\vec{r}}}\ket{\phi_{\vec{k},p}^{(\xi)}}$, which do depend on $\xi$; here, $\vec{K}$ is the position of the $\textrm{K}$ point in the Brillouin zone of the underlying lattice model and $\ket{\phi_{\vec{k},p}^{(\xi)}}$ is the periodic part of the Bloch states, obtained from \equref{eq:Hamiltonian} in our continuum model. The Bloch states encode the non-trivial quantum geometry of the bands and, in particular, that each valley is chiral, with opposite orientations in the two valleys, as required by time-reversal symmetry.

This chirality can, in turn, be illustrated and quantified using the Berry curvature, given by
\begin{equation}
   \Omega_{p}^{(\xi)}{(\bm{\mathrm{k}})} = \left[\nabla_{\bm{\mathrm{k}}} \times \vec{\mathrm{\mathcal{A}}}_{p}^{(\xi)}{(\bm{\mathrm{k}})}\right]_z,
   \label{eq:BerryC}
\end{equation}
where $\vec{\mathrm{\mathcal{A}}}^{(\xi)}_{p}(\bm{\mathrm{k}})=i\braket{\phi^{(\xi)}_{p}{(\bm{\mathrm{k}})}\vert \nabla_{\bm{\mathrm{k}}}\phi^{(\xi)}_{p}{(\bm{\mathrm{k}})}}$ is the Berry connection. As will become important later, we have $\Omega_{p}^{(\xi)}{(\bm{\mathrm{k}})} = \Omega_{p}^{(\xi)}{(-\bm{\mathrm{k}})}$ for our model in \equref{eq:Hamiltonian}. The Berry curvature is sizable only in a ring-shaped region in $\vec{k}$-space, as can be seen in \figref{fig:intro}(c). The resulting valley-resolved Chern number is
\begin{equation}
   C^{(\xi)}_{p} = \frac{1}{2\pi}\int \mathrm{d^2}\bm{\mathrm{k}}\,\Omega^{(\xi)}_{\pm}{(\bm{\mathrm{k}})}= p \xi \frac{n}{2},
   \label{eq:Cnum}
\end{equation}  
showing that it is proportional to the number of layers $n$. The half-integer quantization for odd $n$ is a consequence of the continuum approximation.  

\subsection{Laser field interaction}
Our main goal is to describe the higher harmonics generated by the coupling of the system to a strong pulsed laser with an electric field profile $\bm{\mathrm{E}}(t)=-\partial_t\bm{\mathrm{A}}(t)$. Within the dipole approximation and using the length gauge, this leads to the additional time-dependent term $V_{\text{int}}(t) = -e\bm{\mathrm{E}}(t) \cdot \hat{\vec{r}}$ in the Hamiltonian, i.e., \equref{eq:Hamiltonian} is extended according to $h^{(n)}_{\xi}(\vec{k}) \rightarrow h^{(n)}_{\xi}(\vec{k}) + V_{\text{int}}(t)$. Importantly, the position operator $\hat{\vec{r}}$ acts non-trivially in the Bloch basis, with matrix elements given by
\begin{align}\begin{split}
    \braket{\Psi_{\vec{k},p}^{(\xi)}|\hat{\vec{r}}|\Psi_{\vec{k}',p'}^{(\xi')}} & = \delta_{\xi,\xi'} \bigl[-i \delta_{p,p'} (\nabla_{\vec{k}'} \delta(\vec{k}-\vec{k}')) \\ & \quad + \delta(\vec{k}-\vec{k}') \vec{\mathrm{d}}^{(\xi)}_{pp'}{(\bm{\mathrm{k}})} \bigr],  \label{DipoleMatrixElements}
\end{split}\end{align}
where the dipole matrix elements with respect to the periodic part of the Bloch states are given by
\begin{equation}
   \vec{\mathrm{d}}_{pp'}^{(\xi)}{(\bm{\mathrm{k}})} = i\braket{\phi_{\bm{\mathrm{k}},p}^{(\xi)} \vert \nabla_{\bm{\mathrm{k}}} \phi^{(\xi)}_{\bm{\mathrm{k}},p'}{}}
   \label{eq:DmomEl}
\end{equation}   
and we used that our minimal model in \equref{eq:Hamiltonian} is block diagonal in the valley basis (no intervalley coherence). The diagonal matrix elements are equal to the Berry connection in the respective band and valley, $\vec{\mathrm{d}}_{pp}^{(\xi)}{(\bm{\mathrm{k}})} = \vec{\mathrm{\mathcal{A}}}^{(\xi)}_{p}(\bm{\mathrm{k}})$, probing the chirality of the bands. Importantly, also the off-diagonal components are non-zero which will drive transitions across the band gap. We refer to \appref{sec:AppExp} for the explicit analytical form of the matrix elements that we use in our analysis and here only point out that we use a phase convention where $\vec{\mathrm{d}}_{++}^{(\xi)}{(\bm{\mathrm{k}})} = -\vec{\mathrm{d}}_{--}^{(\xi)}{(\bm{\mathrm{k}})}$ and where $\phi^{(\xi)}_{\bm{\mathrm{k}},p} = \left[\phi^{(-\xi)}_{\bm{\mathrm{k}},p}\right]^*$ such that $\vec{\mathrm{d}}_{pp'}^{(\xi)}{(\bm{\mathrm{k}})} = -\left[\vec{\mathrm{d}}_{pp'}^{(-\xi)}{(\bm{\mathrm{k}})}\right]^*$. Note it always holds $\vec{\mathrm{d}}_{pp'}^{(\xi)}{(\bm{\mathrm{k}})} = \left[\vec{\mathrm{d}}_{p'p}^{(\xi)}{(\bm{\mathrm{k}})}\right]^*$ since the states are normalized.

To solve the time-dependent Schr\"{o}dinger equation $i\ket{\dot{\Psi}(t)}=\hat{H}(t)\ket{\Psi(t)}$, where $\hat{H}(t)$ is the full Hamiltonian including the band structure and the coupling to $\bm{\mathrm{E}}(t)$ as described, we expand the electronic wave function in the Bloch basis, 
\begin{equation}
\ket{\Psi(t)}=\sum_{p=\pm}\int \mathrm{d^2}\bm{\mathrm{k}} \, \alpha_{p,\xi}(\vec{\mathrm{k}},t)\ket{\Psi_{\vec{k},p}^{(\xi)}}.
\end{equation}
As a result of the term in the first line of \equref{DipoleMatrixElements}, the time-dependent Schrödinger equation expressed in terms of the expansion coefficients $\alpha_{p,\xi}(\vec{\mathrm{k}},t) \in \mathbb{C}$ involves momentum derivatives. Translating these coefficients to a ``moving frame'' via $\alpha_{p,\xi}(\vec{\mathrm{k}},t) =: e^{\vec{A}(t)\cdot\nabla_{\vec{K}}} \beta_{p,\xi}(\vec{\mathrm{K}},t)$ with $\bm{\mathrm{K}}=\bm{\mathrm{k}}-\bm{\mathrm{A}}(t)$, eliminates the momentum derivatives. The resultant ``semi-conductor Bloch equations'' (SBEs)~\cite{vampa2017merge,vampa2014theoretical,kim2022theory} then read as
\begin{align}\begin{split}
    \dot{\beta}_{p,\xi}(\vec{\mathrm{K}},t) &= -i \left[ \epsilon_{p}(\vec{k}) + \bm{\mathrm{E}}(t) \cdot \vec{\mathrm{\mathcal{A}}}^{(\xi)}_{p}(\bm{\mathrm{k}}) \right] \beta_{p,\xi}(\vec{\mathrm{K}},t) \\
    &\qquad  -i \bm{\mathrm{E}}(t) \cdot  \sum_{p'\neq p} \vec{\mathrm{d}}_{pp'}^{(\xi)}{(\bm{\mathrm{k}})} \beta_{p',\xi}(\vec{\mathrm{K}},t). \label{SemiconductorBlochEquation}
\end{split}\end{align}
Here and in the following equations, $\vec{k}$ should be understood as $\bm{\mathrm{k}}(t)=\bm{\mathrm{K}}+\bm{\mathrm{A}}(t)$.
Alternatively, this equation can also be derived from the Liouville-von Neumann equation $i\dot{\hat{\rho}}(t)=[\hat{H}(t),\hat{\rho}(t)]$, with associated density matrix $\hat{\rho} = \ket{\Psi(t)}\bra{\Psi(t)}$, which allows to include a dephasing term with associated rate $1/T_2$. The resulting equations can be conveniently stated in terms of the occupations $n_{p,\xi}(\vec{\mathrm{K}},t) = |\beta_{p,\xi}(\vec{\mathrm{K}},t)|^2$ and interband coherence $\pi_{\xi}(\vec{\mathrm{K}},t) = \beta^*_{+,\xi}(\vec{\mathrm{K}},t) \beta_{-,\xi}(\vec{\mathrm{K}},t)$,
\begin{subequations}\begin{align}
    \dot{n}_{p,\xi}(\vec{\mathrm{K}},t) &= i p \bm{\mathrm{E}}(t) \cdot  \vec{\mathrm{d}}_{+-}^{(\xi)}(\vec{k}) \pi_{\xi}(\vec{\mathrm{K}},t) + \text{c.c.}, \\
    \begin{split}\dot{\pi}_{\xi}(\vec{\mathrm{K}},t) &= -i \left[ \Delta\epsilon(\vec{k}) + \bm{\mathrm{E}}(t) \cdot \Delta \vec{\mathrm{\mathcal{A}}}^{(\xi)}(\bm{\mathrm{k}}) - i \frac{1}{T_2} \right] \pi_{\xi}(\vec{\mathrm{K}},t) \\
    & \quad - i \bm{\mathrm{E}}(t) \cdot  \vec{\mathrm{d}}_{+-}^{(\xi)} \Delta n_{\xi}(\vec{\mathrm{K}},t),\end{split}
\end{align}\label{SBESecondForm}\end{subequations}
where $\Delta n_{\xi} = n_{+,\xi} - n_{-,\xi}$ is the occupation imbalance,  $\Delta\epsilon(\vec{k}) = \epsilon_+(\vec{k}) - \epsilon_-(\vec{k})$ the $\vec{k}$-resolved band splitting, and $\Delta \vec{\mathrm{\mathcal{A}}}^{(\xi)}(\bm{\mathrm{k}})=\vec{\mathrm{\mathcal{A}}}^{(\xi)}_{+}(\bm{\mathrm{k}})-\vec{\mathrm{\mathcal{A}}}^{(\xi)}_{-}(\bm{\mathrm{k}})$ is the difference between the Berry connection in the conduction and valence bands.

Throughout this work, we choose the vector potential of the incident laser field to be~\cite{yoshikawa2017high}
\begin{align*}
\bm{\mathrm{A}}(t)=\frac{E_0}{\omega_0\sqrt{1+\epsilon^2}}e^{-(t-t_0)^2/\sigma^2} \left[\cos(\omega_0 (t-t_0))\hat{x} \right.\\  
\left. +\epsilon \sin(\omega_0 (t-t_0))\hat{y}\right]. \numberthis
\label{eq:VecPot}
\end{align*}
Here, $E_0$ is the amplitude of the electric field, $\omega_0$ the frequency, $\sigma^2$ is the variance of the Gaussian envelope, i.e., our measure for the pulse duration, $t_0$ the time at which the pulse reaches its maximum envelope, and $\epsilon$ the ellipticity of the incident electric field. With these conventions, linearly polarized light is defined as $\epsilon=0$, right circular polarization (RCP) corresponds to $\epsilon=-1$ and left circular polarization (LCP) to $\epsilon=1$.

We numerically solve the SBEs in \equref{SBESecondForm} subject to appropriate initial conditions. For instance, at half-filling, we choose $n_{-,\xi}(\vec{K},0)=1$ and $n_{+,\xi}(\vec{K},0)= \pi_{\xi}(\vec{K},0)=0$. We note that in systems with non-trivial topology, the off-diagonal elements of the dipole matrix may have singularities. These must be handled carefully, e.g., by varying the gauge around these points~\cite{chacon2020circular,silva2019high}, enhancing the complexity in solving the SBEs. Although each valley has a non-zero Chern number in our case, this is not an issue for us: as a result of the non-compact nature of the continuum model, we can choose a gauge with perfectly smooth dipole matrix elements, see \appref{sec:AppExp}.

Unless stated otherwise, the specific parameters used in our numerical simulations are as follows: the incident field $E_0=0.005\,$a.u.$=880 \,$GW/cm$^{2}$ (where a.u.~refers to ``atomic units''), with frequency $\omega_0=0.013\,$a.u.=$0.35\,$eV (corresponding to $\lambda=3.5\,\mu$m), $t_0= 25\pi/\omega_0 \approx 146.1\,$fs, $\sigma=59.57\,$fs ($12$ cycles at FWHM), and RMG constants $n=2-6$, $\gamma_1=0.46\,$a.u., bandgap $2w=0.052\,$a.u., $T_2=1.85\,$fs, $\mu=0$ (without doping), $\hbar=\vert e\vert=m_e=k_B=4\pi\epsilon_0=1$. Momenta are measured in units of $k_c=1\,$a.u.. The choice of parameters is discussed further in \appref{MRGParameters}. We use a trapezoidal method for integration over a momentum range $\vert k_i\vert\leq2\,$a.u., $i=x,y$ and a momentum grid of $401\times401$ points. We numerically solve the differential equations in \equref{SBESecondForm} using \emph{Julia ODE solver}.

\subsection{High-harmonic generation}
The interaction with the strong laser field drives the system out of equilibrium, producing a current $\vec{J}(t)$. It is given by the expectation value of the current operator, $\bm{\mathrm{\hat{j}}}=-\bm{\mathrm{\hat{p}}}$ in the gauge we are using here, and can be written as~\cite{yue2022introduction}
\begin{equation}
\bm{\mathrm{J}}(t)=\mathrm{Tr}\{\bm{\mathrm{\hat{j}}}\\,\hat{\rho}(t) \}=i\,\mathrm{Tr}\{ [ \bm{\mathrm{\hat{r}}},\hat{H}(t) ]\,\hat{\rho}(t) \}.
\label{eq:Jobs}
\end{equation}
Recalling that the matrix elements of $\hat{\vec{r}}$ with respect to the Bloch states are given in \equref{DipoleMatrixElements} with intra- and interband contributions, it is natural to also split the current into intra- ($\bm{\mathrm{J}}_{\mathrm{intra}}$) and interband ($\bm{\mathrm{J}}_{\mathrm{inter}}$) terms, $\vec{J} = \sum_\xi (\bm{\mathrm{J}}^{(\xi)}_{\mathrm{intra}} + \bm{\mathrm{J}}^{(\xi)}_{\mathrm{inter}})$. These current contributions can be conveniently written as a function of the Berry curvature, dipole moments, and energy bands as~\cite{golde2008high,haug2009quantum}
\begin{subequations}\begin{equation}
\begin{aligned}
&\bm{\mathrm{J}}^{(\xi)}_{\mathrm{inter}}(t)= \partial_t \bm{\mathrm{P}}^{(\xi)}(t) \\
& \quad = \partial_t \int \diff^2\bm{\mathrm{K}}\,\vec{d}^{(\xi)}_{-+}{(\bm{\mathrm{K}}+\bm{\mathrm{A}}(t))}\pi_\xi(\bm{\mathrm{K}},t) + \text{c.c},
\end{aligned}
\label{eq:Jinter}
\end{equation}
related to the polarization $\bm{\mathrm{P}}^{(\xi)}(t)$ as expressed in the first line, and
\begin{equation}
\bm{\mathrm{J}}^{(\xi)}_{\mathrm{intra}}(t)=\sum_{p}\int \diff^2 \bm{\mathrm{K}} \,\bm{\mathrm{v}}^{(\xi)}_p(\bm{\mathrm{K}}+\vec{A}(t),t) n_{p,\xi} (\bm{\mathrm{K}},t),
\label{eq:Jintra}
\end{equation}\label{TheCurrents}\end{subequations}
respectively. Here, $\bm{\mathrm{v}}^{(\xi)}_p (\bm{\mathrm{k}})= \nabla_{\bm{\mathrm{k}}}\epsilon_{p}(\bm{\mathrm{k}})-\bm{\mathrm{E}}(t)\times \hat{z}\,\Omega_p^{(\xi)} (\bm{\mathrm{k}})$ is the effective velocity, involving both the group velocity associated with the band structure (first term) as well as the anomalous velocity perpendicular to the electric field (second term).

Due to the complex dynamics of the electrons, the associated current $\vec{J}(t)$ not only has frequency components at the fundamental frequency $\omega=\omega_0$ of the incident light, see \equref{eq:VecPot}, but also at higher frequencies, $\omega > \omega_0$, including at higher harmonics, $\omega(l) = l\, \omega_0$ with $l \in \mathbb{N}$, $l>1$. We probe and quantify these aspects using the spectral intensity or Fourier transform of the currents, as explicitly given by the Larmor formula~\cite{yue2022introduction,vampa2014theoretical,ghimire2011observation}
\begin{equation}
I_j(\omega) = \omega^2 \left|\mathrm{FT}_\omega\hspace{-0.2em}\left[ W(t)(\bm{\mathrm{J}}(t))_j\right]\right|^2,
\label{eq:HHG}
\end{equation}
$j=x,y$, where $\mathrm{FT}_\omega[g(t)]$ denotes the Fourier transform of $g(t)$ at frequency $\omega$ and $W(t)$ is the Blackman window function~\cite{vampa2017merge}.

\section{Single-valley limit}\label{SingleValleyLimit}
Having explained the basic setting and formalism used, we can now discuss our findings. Our primary goal is to study valley-polarized RMG, where interaction effects spontaneously create an asymmetry in the two valleys. To keep the discussion transparent, we therefore assume an extreme anisotropy first, where only one of the two valleys needs to be taken into account in the expressions above, while the other one will remain ``inactive''. In \secref{IncludingSecondValley} below, we will then study the changes when both valleys with different gaps are included simultaneously. Without loss of generality, we will take the $\xi=+$ valley as ``active'', but suppress the superscript $\xi=+$ in the following expressions for notational simplicity.

\begin{figure}[tb]
    \centering
    \includegraphics[width=\columnwidth]{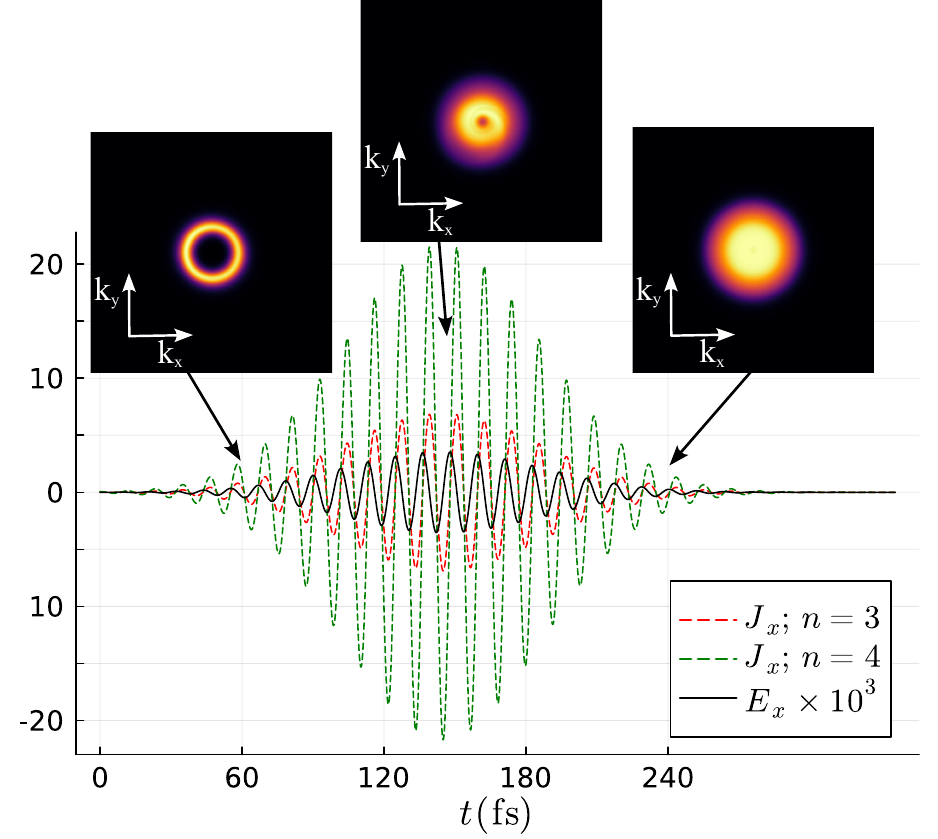}
    \caption{Time dependence of the $x$ component of the total current $\vec{J}(t)=\vec{J}_{\mathrm{inter}}(t)+\vec{J}_{\mathrm{intra}}(t)$, for $n=3$ (red), $4$ (green), both with incident light with RCP. The electric field in the $x$ direction is shown in black. The three insets display the momentum dependence of the population for $n=4$ of the conduction band at three different times as indicated by the black arrows; specifically, left panel at $60.5\,$fs with maximum population (yellow) of $\mathrm{max}=0.0045$, middle panel ($\mathrm{max}=0.38$) at $133\,$fs,  right panel ($\mathrm{max}=0.496$) at $241.8\,$fs.}
    \label{fig:Current}
\end{figure}

Before studying the non-perturbative regime and higher harmonics, let us first gain some intuition by looking at the behavior right at the beginning of the pulse. Starting from the filled-lower band initial condition, we can neglect in the early-time limit of \equref{SemiconductorBlochEquation} the finite population in the conduction band that will eventually be induced due to the interaction with light over a finite time period. Then the resulting equation for $\beta_+(\vec{k})$ can be written as
\begin{equation}
    \partial_t \beta_{+}(\vec{k}) = - i\,\bm{\mathrm{E}}(t)\cdot \vec{\mathrm{d}}_{+-}(\bm{\mathrm{K}}+\bm{\mathrm{A}}(t)) + \mathcal{O}(\beta_+).
\end{equation}
So we expect that the population $n_{+}(\bm{\mathrm{K}},t)$ in the upper band initially grows the strongest in momentum space where the interband matrix elements of $\vec{\mathrm{d}}$ are the largest. Inspection of \figref{fig:intro}(d) shows that this quantity is peaked, similar to the Berry curvature, on a ring of finite radius. As can be seen in the (leftmost) inset of \figref{fig:Current}, this is indeed where we also find the largest initial occupation in the conduction band right at the beginning of the pulse. So the momentum-space distribution is primarily determined by the Bloch-state wave functions rather than the band energies, which are still fairly flat in this momentum range. 

\subsection{High harmonic generation}

In the main panel of \figref{fig:Current}, we can see that this non-equilibrium configuration gives rise to an oscillatory current. It is consistently synchronized with a $\pi/2$ phase delay with the incident laser field, in line with expectations based on previous numerics in different systems~\cite{bera2023topological}; see also \appref{sec:AppCurCircular} for further discussion. Interestingly, we further observe that the current magnitude increases with $n$, which in our effective two-band model description simply determines the power of the complex momentum factor $k_\xi$ in \equref{eq:Hamiltonian}. Note we only show $n=3,4$ for clarity but the trend continues, which we have checked for $n=2,5,6$. We have verified, by artificially keeping $\epsilon_{p}(\vec{k})$ fixed and only varying the Bloch states with $n$, that the $\ket{\phi_{p}{(\bm{\mathrm{k}})}}$, entering the transition matrix elements, and thus the quantum geometry provide the dominant role in this enhancement. This further agrees with the naive expectations based on the early-time analysis above.

\begin{figure*}[tb]
    \centering
    \includegraphics[width=2\columnwidth]{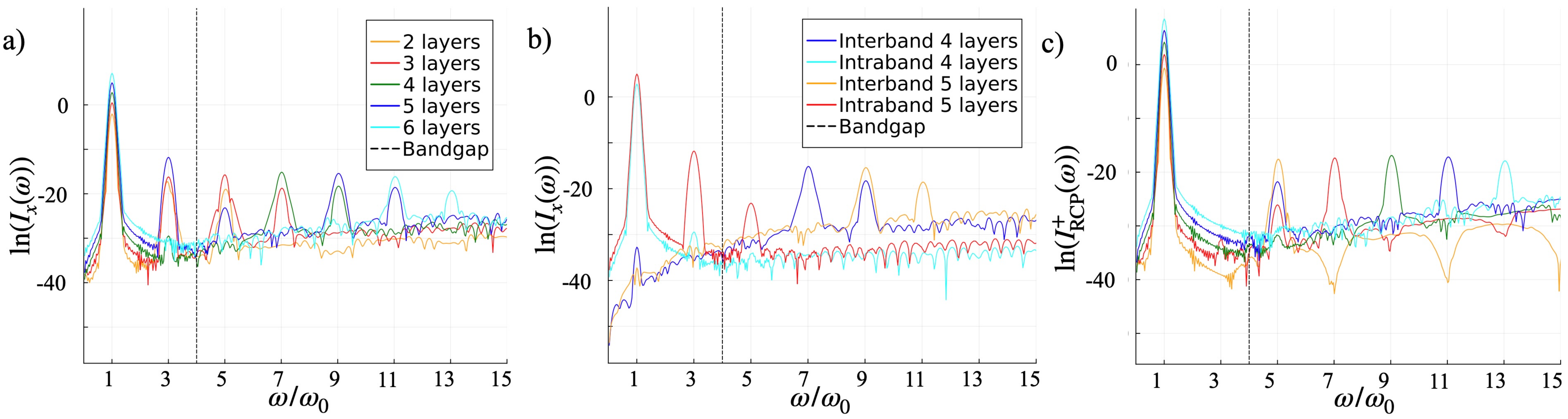}
    \caption{a) HHG spectrum $\mathrm{ln}[I_x(\omega)]$, see \equref{eq:HHG}, for $n=2-6$. b) $\mathrm{ln}[I_x(\omega)]$ split into interband and intraband contributions for $n=4$ and $5$. c) $\mathrm{ln}[I^+_{\mathrm{RCP}}(\omega)]$, see \equref{IpmIntensity}, for $n=2-6$, where the same legend as in (a) applies. The input laser has RCP in all cases. Dashed black line shows the bandgap.} 
    \label{fig:HHG}
\end{figure*}

To gain more information, we use \equref{eq:HHG} to perform a spectral decomposition and analyze the higher harmonics. The result is displayed in \figref{fig:HHG}(a), where we measure $\omega$ in units of $\omega_0$ such that the harmonics with frequency $w(l)$ correspond to the integer values $l$. We observe prominent odd harmonic orders, while even $l$ are suppressed. As we explain in more detail in \appref{OnlyCertainHarmonics}, one can understand this behavior intuitively (within the Keldysh approximation) to be the result of $\Omega_{p}{(\bm{\mathrm{k}})} = \Omega_{p}{(-\bm{\mathrm{k}})}$ and $\vec{d}_{+-}(\vec{k}) =-(-1)^n \vec{d}_{+-}(-\vec{k})$, i.e., the definite parity of these matrix elements in $\vec{k}$. Among the odd values of $l$, the dominant harmonic above the band gap scales with $n$ like $l=2n-1$ (and $l=2n+1$ being the second largest; in \appref{OnlyCertainHarmonics} we also provide a qualitative argument for why specifically those harmonics dominate). This shows that the degree of winding of the Bloch states is directly connected to the dominant harmonic. 

It is further interesting to investigate the intra- and interband currents separately. As can be seen in \figref{fig:HHG}(b), where we focus on $n=4,5$ for clarity (but we checked that the following also holds for $n=2,3,6$), the strong incident laser field and small bandgap favor the interband contribution in higher harmonic orders, above the bandgap. More specifically, the interband HHG is dominant for $l\geq2n-1$. At the same time, we can see that intraband HHG shows different signatures for even and odd $n$: For odd $n$ the intraband is dominant up to the $n$th harmonic order (HO), while for even $n$, it only contributes significantly to the first HO.
Since $\bm{\mathrm{v}}_p(\bm{\mathrm{k}})$ in \equref{eq:Jintra} does not depend on the parity of $n$, the difference in the number of HOs can only come from how the bands are populated. 

For HOs above $2n+1$, the HHG spectrum does not show any pronounced peaks, while for higher HOs (for example, above $60\,\omega_0$ for $n=4$ as shown in \figref{fig:HHGAll} in \appref{sec:AppHHG}) clearer peaks reemerge. This result resembles the HHG in \cite{boyero2022non} for single-layer graphene.

\subsection{Circular dichroism}
\label{sec:CD}
To provide a more direct link to the chiral nature of the single-valley band structure, we next address circular dichroism (CD) \cite{asteria2019measuring,ranjbar2009circular,chacon2020circular,kim2022theory}, i.e., contrast the response of the system to light with RCP and LCP. To quantify it for the different HOs, we employ the commonly used dimensionless quantity
\begin{equation}
\mathrm{CD}_l=\frac{I^+_{\mathrm{RCP}}(l)-I^-_\mathrm{LCP}(l)}{I^+_{\mathrm{RCP}}(l)+I^-_{\mathrm{LCP}}(l)}, 
\label{eq:CD}
\end{equation}
where $l$ is the HO, as before associated with the frequency $\omega(l)=l\omega_0$, and  
\begin{equation}
    I^\pm_P(l)=\omega^2(l) \left|\mathrm{FT}_{\omega(l)}[ W(t)(J_x(t)\pm iJ_y(t))]\right|^2  \label{IpmIntensity}
\end{equation}
with polarization $P \in \text{RCP}, \text{LCP}$ of the incoming light. Intuitively, $I^+_{\mathrm{RCP}}$ ($I^-_\mathrm{LCP}$) measures the right-handed (left-handed) response to a right-handed (left-handed) perturbation; as such, a finite difference $I^+_{\mathrm{RCP}}(l)-I^-_\mathrm{LCP}(l)$ implies that the system has a preferred chirality and the denominator in \equref{eq:CD} normalizes $\mathrm{CD}_l$ to become a dimensionless quantity.

However, before studying $\mathrm{CD}_l$, let us first take a closer look at $I^+_{\mathrm{RCP}}$ as an example, which is represented in Fig.~\ref{fig:HHG}(c) for $n=2-6$. We observe a dominant peak at HO $l=2n+1$; as expected, $I^-_{\mathrm{LCP}}$ displays similar behavior (not shown). In contrast, for the orthogonal polarizations $I^+_{\mathrm{LCP}}$ and $I^-_{\mathrm{RCP}}$, we obtain a dominant peak at HO $2n-1$ instead (see Fig.~\ref{fig:ImRCP} in \appref{sec:AppHHG}). 
Writing the $x$-component of the current as the linear combination $J_x=\frac{1}{2}(J^++J^-)$, with $J^\pm=J_x\pm iJ_y$, it becomes clearer how $I_x$ contains the HOs of both $I^\pm$, which is consistent with \figref{fig:HHG}(a), displaying peaks at both HOs $l=2n\pm1$. 

Just as before, the harmonics above the band gap are dominated by the interband current, while the peaks below the bandgap are due to the intraband contribution.  

In the inset of \figref{fig:CD}(a), we show $\text{CD}_{l}$ for the aforementioned dominant harmonic $l=2n+1$. It is not only clearly non-zero and thus reflects the chirality of the system, but it is also very close to its maximum value of $1$, i.e., $I^+_{\mathrm{RCP}} \gg I^-_\mathrm{LCP}$, in particular for $n \geq 3$. For the other valley, $\text{CD}_{l}$ will have the same magnitude but opposite sign. By virtue of saturating the bound $|\text{CD}_{2n+1}| \leq 1$, this quantity is not sensitive to the number of layers $n$. Inspired by \refcite{silva2019topological}, which proposed to compare the \textit{sign} of $\text{CD}_{1}$ with $\text{CD}_{l}$ for higher $l$ above the gap to extract the topological properties, we here consider their ratio, $c_n = \text{CD}_{2n+1}/\text{CD}_{1}$. We find $c_n < 0$ in line with the expectations for a topological regime of a two-band model \cite{silva2019topological}. This is also consistent with our observation in \appref{sec:AppCurCircular} that the phase relation between the $x$ and $y$ components of the current is opposite for the inter- and intraband current contributions.
What is more, our numerics reveals that the magnitude $|c_n|$ does depend sensitively on $n$: as can be seen in \figref{fig:CD}(a), $\ln |c_n|$ increases approximately linearly with $n$ and, thus, constitutes a natural quantifier of the degree of chirality in our model. 

\subsection{The effect of doping}
\label{sec:Doping}
We next address what happens once the system is doped away from half filling, which in RMG is done in experiment by the application of gate voltages and, in fact, is even needed for the spontaneous emergence of valley imbalance \cite{valleySCExp}. In our calculation, this implies that different initial conditions~\cite{dutta2025probing} have to be used when solving the SBEs in \equref{SBESecondForm}. Specifically, we will impose $\pi(\vec{K},0) = 0$ and $n_p(\vec{K},0) = n_{\text{F}}(\epsilon_p(\vec{K})-\mu)$, $n_{\text{F}}(\epsilon)=\frac{1}{e^{\epsilon/T}+1}$, with a suitably chosen chemical potential $\mu$; our previous half-filled initial conditions correspond to $\mu=0$ (and $T\rightarrow 0$). Away from half-filling, we use a small but finite $T$ of $T= 0.01\,\text{a.u.}$.

\begin{figure}[tb]
    \centering
    \includegraphics[width=\columnwidth]{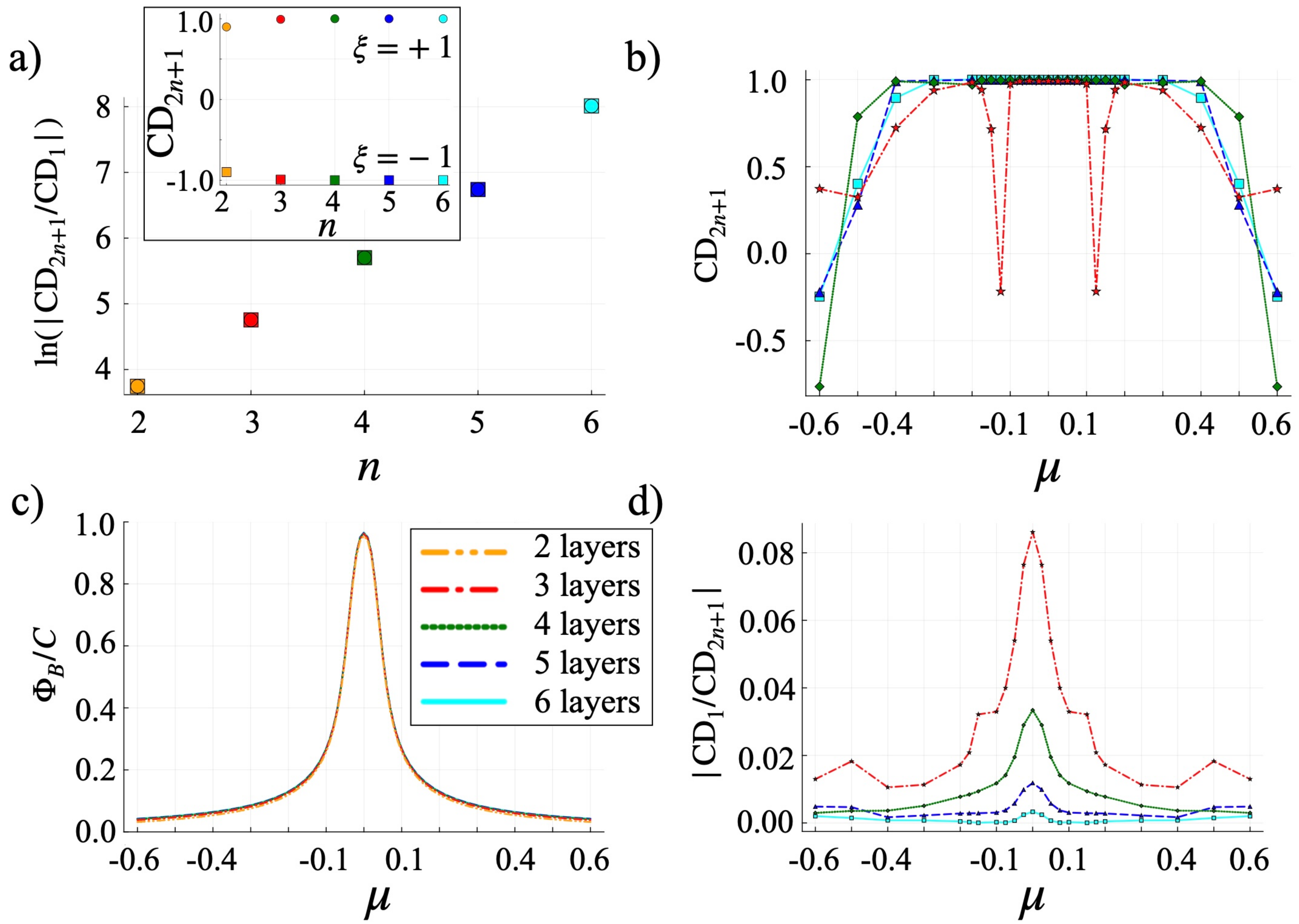}
    \caption{a) $\mathrm{ln}(\vert c_n\vert)=\mathrm{ln}(\vert\mathrm{CD}_{2n+1}/\mathrm{CD}_{1}\vert)$, see \equref{eq:CD}, as a function of the number of layers for $n=2-6$. The result for $\xi=-$ (squares) and $\xi=+$ (circles, not visible) is identical. The inset shows the value of $\mathrm{CD}_{2n+1}$ as a function of $n$ for $\xi=\pm$.  b) Value of $\mathrm{CD}_{2n+1}$ for $n=3-6$ as a function of the chemical potential; $\mathrm{CD}\sim 1$ in the region $\mu<16w$, for $n=4-6$. c) $\Phi_B(\mu)/C$, see \equref{PhiBDefinition}, as a function of the chemical potential $\mu$ for $n=2-6$. d) Relative helicity $1/\vert c_n\vert=\vert\mathrm{CD}_{1}/\mathrm{CD}_{2n+1}\vert$ in presence of doping. It shows a peak in the region $\mu<4w$, similar to $\Phi_B(\mu)/C$. Note that the data point for $n=3$ with $\mu=0.125$ was removed [cf.~sharp dip in part (b)] for better visualization of the results. Legend of all figures in (c) and $\mu$ in a.u.~in all plots.}
    \label{fig:CD}
\end{figure}

The key features of our results remain qualitatively intact when detuning $\mu$ slightly away from half-filling, despite the large density of states associated with the flat band bottom. For instance, in \figref{fig:CD}(b) we display the $\mu$ dependence of $\text{CD}_{2n+1}$, which stays constant and close to saturation at the bound in a finite (and, as expected, particle-hole symmetric) region around $\mu=0$ for all $n$. In particular, for $n \geq 4$, this region is fairly large---much larger than the region where the net Berry curvature, 
\begin{equation}
    \Phi_B(\mu)= \frac{1}{2\pi} \sum_{p}\int\diff^2\vec{k}\,\Omega_{p}(\bm{\mathrm{k}})n_{\text{F}}(\varepsilon_{p}(\bm{\mathrm{k}})), \label{PhiBDefinition}
\end{equation}
of the occupied states is maximal and close to the quantized value in \equref{eq:Cnum}, see \figref{fig:CD}(c). Similar to the $n$ dependence at half-filling, we expect that the weak sensitivity of $\text{CD}_{2n+1}$ to detuning $\mu$ from zero compared to $\Phi_B$ is due to the fact that $\text{CD}_{2n+1}$ is saturating the bound. Instead, $c_n$ is expected to also vary more strongly with $\mu$, which is indeed confirmed by \figref{fig:CD}(d); we can see that $1/|c_n| = |\text{CD}_{1}/\text{CD}_{2n+1}|$ exhibits a peak in the region $\mu<\vert 4w\vert$, similar to $\Phi_B$. The decrease of $|\text{CD}_{1}/\text{CD}_{2n+1}|$ with doping is naturally rooted in the decay of $\text{CD}_{1}$ with doping: as $\text{CD}_{1}$ is dominated by intraband current contributions it is crucially determined by the Berry curvature of the occupied states \cite{silva2019high} and, hence, should qualitatively follow $\Phi_B(\mu)$ in \equref{PhiBDefinition}. 

\section{Including the second valley}\label{IncludingSecondValley}

\begin{figure}[tb]
    \centering
    \includegraphics[width=\columnwidth]{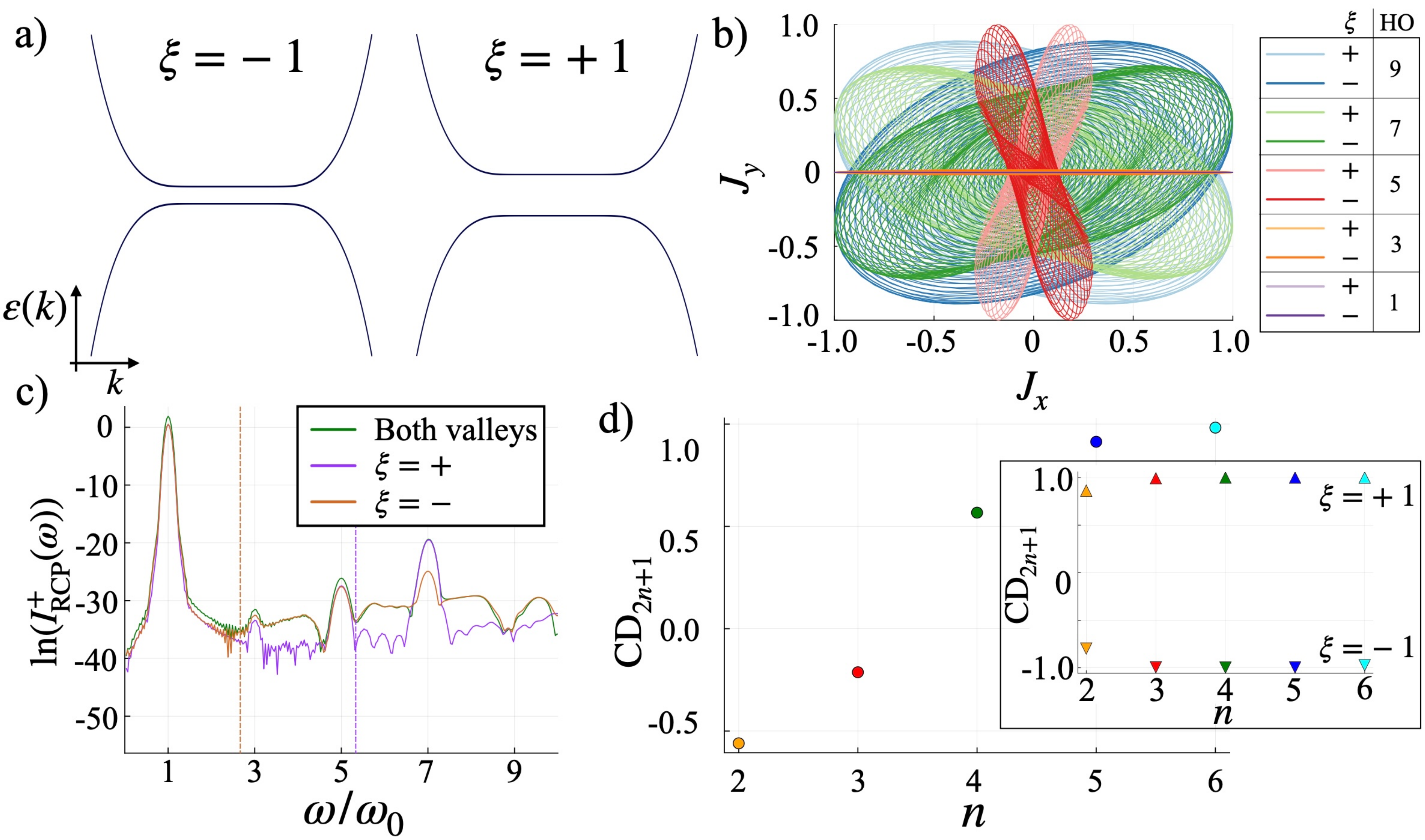}
    \caption{a) Schematic illustration of the electronic dispersion relation in the two valleys in the presence of valley imbalance, with smaller (larger) gap for $\xi=-$ ($\xi=+$). b) Normalized time trace of currents for the first $9$ HOs for valley $\xi=+$ (dark colors) and $\xi=-$ (light colors). Here $n=4$ and the incident laser is linearly polarized along the $x$ direction. c)~$\mathrm{ln}[I^{+}_{\mathrm{RCP}}(\omega)]$ for $n=3$ as function of $\omega/\omega_0$ when both valleys are simultaneously active (green) and for the two valleys separately (purple and orange). d) $\mathrm{CD}_{2n+1}$ including both valleys. Inset shows value of $\mathrm{CD}_{2n+1}$ in each valley, upper triangles correspond to $\xi=+$ and lower triangles to $\xi=-$. Valley $\xi=+$ has larger gap $2(w +\Phi_V)=8w/3$ and valley $\xi=-$ has smaller gap $2(w -\Phi_V)=4w/3$; $\Phi_V=w/3$.}
    \label{fig:2VP}
\end{figure}

After clarifying the HHG spectrum in a single-valley, we finally address the case where both valleys are taken into account. To incorporate the interaction-induced valley polarization in the system, we follow previous works (see, e.g., \cite{StandfordWithTheRelativeChirality,Paco,YahuisPaper,OurEnergetics,2025arXiv251019943S,FranzSCRingOfFire,PatrickLeeControl,BerryTrashcan}) and employ a (mean-field) valley imbalance order parameter $\Phi_V$. While the Bloch states are unaffected by it, the band energies in \equref{eq:Energy} are modified according to $\epsilon_{\pm}(\vec{k}) \rightarrow \mathcal{E}_{\pm}(\vec{k},\xi) = \varepsilon_{\pm}(\vec{k}) \pm \xi\,\Phi_V$, which modifies the bandgap in valley $\xi$ according to $2w \rightarrow 2(w + \xi \,\Phi_V)$, see the schematic dispersion relation illustrated in \figref{fig:2VP}(a). 

In our low-energy continuum approximation, the valley effectively acts as a quantum number, and the contributions to the currents from the two valleys simply add up, as reflected by the expressions in \secref{ModelAndApproach}, in particular, \equref{TheCurrents}. To quantify the HHG, we still use the Larmor expression in \equref{eq:HHG} but replace $\vec{J}$ with the average instead of the sum of the currents in two valleys, i.e., $\vec{J} = \frac{1}{2}\sum_\xi (\bm{\mathrm{J}}^{(\xi)}_{\mathrm{intra}} + \bm{\mathrm{J}}^{(\xi)}_{\mathrm{inter}})$, to be able to better compare with the single-valley limit; the associated spectral intensity will be denoted by $I^{(2)}_j(\omega)$ in the following. Since we have seen above that the main features of the HHG are, to a good approximation, unaffected when varying $\mu$ in a finite range around zero, we here focus on half filling ($\mu=0$) for concreteness. 

As a first illustration of the contrasting behavior and interference of the two valleys concerning the generation of currents, we consider linearly polarized light, with the electric field along the $x$ direction. We Fourier transform the time-dependent current distribution, apply a Gaussian mask around each HO, followed by a Fourier transform back to the time domain (a similar approach was used in \cite{PhysRevB.110.054103}). The resulting time traces of the current $\vec{J}$ are shown in \figref{fig:2VP}(b) for a few selected harmonics and studying the two valleys separately. 
First, we can see how the polarizations of low HOs are aligned with the laser field. Meanwhile, above the first band gap, the anomalous current in the $y$ direction becomes significant. This is a direct manifestation of the low symmetries of the system, which does not exhibit a reflection symmetry with respect to the $xz$-plane. However, as a result of the reflection at the $yz$-plane which relates the two valleys, we can see that the two traces of the two valleys are mirror images of one another. If both valleys are taken into account simultaneously, we find that the $y$ component in the time traces of the current cancel out to a very good degree and the current is approximately along the direction of the electric field. 

To address the physics of the two valley system, we therefore consider circularly polarized light. In \figref{fig:2VP}(c), we show the HHG spectrum, $I^{(2)}(\omega)$, for RCP of the incident light in the two-valley system, where the gap in one valley is larger (in the example $\xi=+$) than in the other ($\xi=-$). Direct comparison with the analogous single-valley plot in \figref{fig:HHG}(c) [and the other single-valley curves in \figref{fig:2VP}(c)] shows that the central features---the prominent harmonics, particularly at $l=2n \pm 1$---are qualitatively unaffected, as the wavefunctions remain intact and interference effects of the currents between the two valleys do not qualitatively change these features. 

Closer inspection of \figref{fig:2VP}(c) shows that the dominant peak in $I^{(2)}$ away from the fundamental harmonic appears at the HO ($2n+1$) where also the single valley $\xi=+$ has its dominant peak with virtually identical peak height as the two-valley system. We checked that this is also the case for $n=4,5,6$ (for $n=2$, the associated HO is below the band gap of the larger valley). To understand whether it is the difference in band gap or chirality that selects the $\xi=+$ valley to determine the peak of the two-valley problem, we investigate the behavior under LCP light while keeping the band gaps identical: now, it is the $\xi=-$ valley that drives the dominant higher harmonic peak of $I^{(2)}(\omega)$. In line with other recent results \cite{mrudul2021controlling,heras2026pulse} and as expected by symmetry, this shows that circularly polarized light couples preferably to band structures with a specific chirality. 

The two valleys have opposite chirality [cf.~\equref{eq:Cnum}] and, thus, opposite $\mathrm{CD}_l$ by themselves [see also inset in \figref{fig:2VP}(d)]. Therefore, it is natural to expect a more non-trivial interplay of the two valleys in CD. This is indeed what we find, as shown in the main panel of \figref{fig:2VP}(d): although the gaps are fixed (in particular, in all cases $\text{CD}_{2n+1}$ of the single-valley problem is negative in the valley, $\xi=-$, with the smaller gap), we find that $\text{CD}_{2n+1}$ changes sign as a function of $n$. This shows that, while $\text{CD}_{2n+1}$ is dominated for $n\leq3$ by the valley with smaller gap, it is the larger-gap valley which determines the sign of $\text{CD}_{2n+1}$ for $n \geq 4$. Notably, at $n=3$, we have a near cancellation between the opposite chiralities of the two valleys. For $n=4-6$, $\mathrm{CD}_{2n+1}$ is more and more determined by the larger-gap valley ($\xi=+$) upon increasing $n$, culminating in $n=6$, where $\mathrm{CD}_{2n+1}$ is close to reaching the maximum value of $+1$.

\section{Conclusion}\label{Conclusion}
In this work, we studied harmonic generation due to a high-intensity light pulse in a minimal model for rhombohedral multi-layer graphene. We investigated the behavior as a function of the number $n$ of graphene layers, explored the dependence on the band filling and the interplay of two energetically split valleys, scrutinizing the role of the quantum geometry of the chiral Bloch states.

We first started by investigating the physics of a single valley. We showed that the momentum-space occupations at early times after the onset of the light pulse are crucially shaped by the non-trivial momentum dependence of the Bloch states, which has the characteristic feature of being significant in a ring of finite radius around the $\mathrm{K}$ and $\mathrm{K}'$ points. Since the Bloch state's winding in $\vec{k}$ space crucially affects the transition matrix elements, it is natural to have a huge impact on the HHG spectra: The dominant harmonic orders above the bandgap scale linearly with $n$ and are primarily driven by the interband contribution to the current. More specifically, we find that the dominant peaks of the spectral intensity in \equref{eq:HHG} are at frequencies $\omega(l) = l \omega_0$ with $l=2n\pm 1$; we only find one of these two peaks in the analogous intensity $I^{\pm}_P$ in \equref{IpmIntensity}, depending on $\pm$ and the chirality of the polarization $P$.

To probe the chirality of the valleys systematically, we investigated
circular dichroism of the different harmonic orders $l$, specifically, the dimensionless quantity in \equref{eq:CD}. While it was found to be almost exclusively reaching its maximum magnitude of one, its sign is determined by the chirality of the normal state, i.e., which of the two valleys dominates. We identified the quantity $\text{CD}_{1}/\text{CD}_{2n+1}$ to also provide quantitative information about the \textit{degree} of the winding of the Bloch states since $\ln |\text{CD}_{1}/\text{CD}_{2n+1}|$ is found to scale linearly with $n$.

We further studied the doping dependence, which revealed that $\text{CD}_{2n+1}$ stays close to being saturated in an extended range of chemical potential values, much larger than the gap and the range where the net Berry curvature of the occupied states, $\Phi_B$ in \equref{PhiBDefinition}, stays close to its value at charge neutrality (the Chern number). It is again the normalized quantity $\text{CD}_{1}/\text{CD}_{2n+1}$ that follows $\Phi_B$ more closely. Finally, we investigated the interplay of two opposite chiralities when both valleys, but with different interaction-induced gaps, are included. Remarkably, the dominant valley determining the sign of $\text{CD}_{2n+1}$ depends on the number of layers $n$.

Taken together, our findings show that rhombohedral multilayer graphene, particularly with valley imbalance, and chiral multi-band systems in general display rich physics concerning harmonic generation and circular dichroism. As we discussed in detail, this is rooted in their non-trivial quantum geometry. Given the fast progress in non-linear optics in two-dimensional materials \cite{Review2DHG,AnotherReview2D,zhou2022engineering}, we hope that some of our predictions will become relevant to future experiments in multi-layer graphene and related chiral systems.  

\begin{acknowledgments}
J.O.d.A. and M.S.S. acknowledge support from EU HORIZON-MSCA-2024-PF-01 Project No.$101205617$, GATOR.
W.K.-K. acknowledges the Laboratory Directed Research and Development (LDRD) program of Los Alamos National Laboratory under projects number 20260116ER and 20240037DR.
\end{acknowledgments}

\bibliography{draft_Refs}

\onecolumngrid

\begin{appendix}

\section{Analytical expressions for wavefunctions and matrix elements}
\label{sec:AppExp}

Decomposing \equref{eq:Hamiltonian} in the Pauli basis as $h_{n}(k)=\mathrm{g}_0\sigma_0+\bm{\mathrm{g}}(k) . \bm{\sigma}$, where $\bm{\sigma}$ are the Pauli matrices and $\bm{\mathrm{g}}(k) = (\mathrm{g}_1(k),\mathrm{g}_2(k),\mathrm{g}_3(k))$ are
\begin{align*}
&\mathrm{g}_0=-\mu \hspace{1.6cm} \mathrm{g}_1(k)=(-1)^{n+1}\gamma_1\frac{k^{n}}{k_c^{n}}\cos{(n\xi\beta)}\\ 
&\mathrm{g}_2(k)=(-1)^{n+1}\gamma_1\frac{k^{n}}{k_c^{n}}\sin{(n\xi\beta)} \hspace{0.9cm} \mathrm{g}_3(k)=w \numberthis
\end{align*}
where $k=\sqrt{k_x^2+k_y^2}$ and $\beta=\arctan\left(k_y/k_x\right)$.

The eigenvectors of the Hamiltonian in \equref{eq:Hamiltonian} are
\begin{equation}
    \ket{\phi^{(\xi)}_{-}(k)}= \frac{e^{i n \xi\beta}}{ \sqrt{2 k^{2n} \vert\bm{\mathrm{g}}(k)\vert \left(\mathrm{g}_3(k) - \vert\bm{\mathrm{g}} (k)\vert\right)} }
    	\left( \begin{matrix}
   
                                       \mathrm{g}_3(k) - \vert\bm{\mathrm{g}}(k)\vert \\
                                       -\gamma_1 k^{n} e^{-i n \xi\beta}/k^n_c
                                      \end{matrix}	
                                \right),
          \label{eq:Evec1}
\end{equation}   

\begin{equation}
    \ket{\phi^{(\xi)}_{+}(k)}= \frac{1}{\sqrt{2 \vert\bm{\mathrm{g}}(k)\vert\left(\mathrm{g}_3(k) + \vert\bm{\mathrm{g}}(k)\vert\right)}} 
    	\left( \begin{matrix}
   
                                       \mathrm{g}_3(k) + \vert\bm{\mathrm{g}}(k)\vert \\
                                      -\gamma_1 k^{n} e^{-i n \xi\beta}/k^n_c
                                      \end{matrix}	
                                \right),
                \label{eq:Evec2}
\end{equation}   
where $\vert\bm{\mathrm{g}}(k)\vert=\sqrt{(\mathrm{g}_1(k))^2 + (\mathrm{g}_2(k))^2 + (\mathrm{g}_3(k))^2}$ and $k e^{i n \xi\beta}= (k_x + i \xi k_y)^{n}$.

The dipole moment matrix $\vec{\mathrm{d}}^{(\xi)}(\vec{k})$ is
\begin{equation}
   \vec{\mathrm{d}}^{(\xi)}(\vec{k})=\left(
  \begin{matrix} 
   
                                       d_{--}{(\bm{\mathrm{k}})} & d_{-+}{(\bm{\mathrm{k}})} \\
                                       d_{+-}{(\bm{\mathrm{k}})} & d_{++}{(\bm{\mathrm{k}})}
                                      \end{matrix}	 \right)                                               
\end{equation}   

with exact solutions to \equref{eq:DmomEl} as
\begin{equation}
d_{--}^{(\xi)}{(\bm{\mathrm{k}})}=\left(\frac{-\xi k_y\, n \left( w - \sqrt{\,w^{2} + k_c^{-2n} (k_x^{2} + k_y^{2})^{n} \gamma^{2}} \right)}{ 2 (k_x^{2} + k_y^{2}) \sqrt{\,w^{2} + k_c^{-2n} (k_x^{2} + k_y^{2})^{n} \gamma^{2}}}\right)\hat{x}+\left( \frac{ -\xi k_x\, n \left( -w + \sqrt{\,w^{2} + k_c^{-2n} (k_x^{2} + k_y^{2})^{n} \gamma^{2}} \right)}{2 (k_x^{2} + k_y^{2}) \sqrt{\,w^{2} + k_c^{-2n} (k_x^{2} + k_y^{2})^{n} \gamma^{2}}}\right)\hat{y}
\end{equation}

\begin{equation}
\begin{aligned}
d_{+-}^{(\xi)}{(\bm{\mathrm{k}})}=&\left(\frac{
 i \,\xi \bigl( -k_c (k_x + i \xi k_y) \bigr)^{n} \, n \gamma  \left( i \xi k_x w + k_y \sqrt{\,w^{2} + k_c^{-2n} (k_x^{2} + k_y^{2})^{n} \gamma^{2}} \right)}{2 (k_x^{2} + k_y^{2}) \left( k_c^{2n} w^{2} + (k_x^{2} + k_y^{2})^{n} \gamma^{2} \right)} \right)\hat{x} \,+ \\ 
& +\left( -\frac{ i \xi\, \bigl( -k_c (k_x + i\xi k_y) \bigr)^{n} \, n \gamma \left( - i \xi k_y w + k_x \sqrt{\,w^{2} + k_c^{-2n} (k_x^{2} + k_y^{2})^{n} \gamma^{2}} \right)}{ 2 (k_x^{2} + k_y^{2}) \left( k_c^{2n} w^{2} + (k_x^{2} + k_y^{2})^{n} \gamma^{2} \right)}\right)\hat{y}
\label{d+-}
 \end{aligned}
\end{equation}

\begin{equation}
\Omega^{(\xi)}_{\pm}(\bm{\mathrm{k}})= \pm\frac{\xi n^2 \, w\, \gamma^2\, k_c^{-2n}(k_x^2+k_y^2)^{n-1}}{2(w^2+\gamma^2 \, k_c^{-2n} \,(k_x^2+k_y^2)^{n})^{3/2}}
\end{equation}

Analyzing the analytical expressions, we can derive the solutions  $\partial_k \varepsilon_\pm (k)\propto \pm n k^{2n-1} a/(b+ k^{2n})^{1/2}$ and Berry curvature $\Omega_{\pm}(k)\propto \pm c \,n^2 k^{2n-2}/(b+k^{2n})^{3/2}$, ($a=\gamma^2$, $b= w^2 /\gamma^2$,  $c= w \gamma^2/2$ ), contribute to the intraband current, the phase relation between the $\hat{x}$ and $\hat{y}$ components exchanging and rotating the $k_x$ and $k_y$ components of the dipole matrix elements are $d^{(\xi)}_{+-}(k_y,-k_x)\hat{x}=e^{-i n\pi/2}d^{(\xi)}_{+-}(k_x,k_y)\hat{y}$ and $d^{(\xi)}_{--}(k_y,-k_x)\hat{x}=d^{(\xi)}_{--}(k_x,k_y)\hat{y}$.  

\section{Rhombohedral graphene parameters}
\label{MRGParameters}
Here, we comment on the relation of the parameters used in the main text and those believed to describe RMG well. As for the latter, the interlayer nearest neighbor hopping amplitude between superposed sites $B_{j}$-$A_{j+1}$ is $\gamma_1=0.381\,$eV$=1.4\times10^{-2}\,$a.u., the intralayer nearest neighbor $\gamma_0=3.16\,$eV$=0.12\,$a.u., and the intralayer distance is $a=0.246\,$nm$=4.65\,$a.u.~\cite{muten2021exchange}, yielding a momentum scale of $k_c=2\gamma_1/(\sqrt{3}a\gamma_0)\sim0.03\,$a.u.. The bandgap ($2w$) can be up to $42\,$meV $=1.54\times10^{-3}$a.u. as implemented in~\cite{lee2014competition}. In our numerical simulations, the relevant parameters $k_c$, $\gamma_1$ and $2w$ are, respectively chosen as $1\,$a.u., $0.46\,$a.u. and $0.052\,$a.u.. These parameters differ from the true values by a constant factor of $\sim33.33$; it can be thought of as a rescaled choice of units such that the momentum scale $k_c$ is fixed to unity, $k_c\sim1\,$a.u.. 

However, transforming back from the parameters used in the simulations to those of RMG listed above is straightforward. Most importantly,  the ratio between the bandgap and the laser frequency of $2w=4\hbar\omega_0$ for rhombohedral graphene with a bandgap of $42\,$meV would require a laser with frequency $\omega_0=2.5\,$THz ($\lambda=124\,\mu$m) which is possible with current technology~\cite{hafez2018extremely}.

\section{Keldysh approximation and semiclassical analysis}
\label{OnlyCertainHarmonics}

\subsection{Why only odd harmonics}

In this appendix, we would like to qualitatively understand the dominance of odd harmonics above the bandgap in the HHG. To do that, we focus on the interband current, as it is the dominant contribution above the bandgap. Using the Keldysh approximation~\cite{yue2022introduction,vampa2014theoretical,li2023high}, we will explore the parity of the matrix elements under inversion of momentum, $\vec{k}\rightarrow-\vec{k}$, and the periodicity of the incident field to identify the dominant HOs.

The Keldysh approximation fixes the difference between the band populations to $1$, i.e. $n_{+,\xi}(\vec{\mathrm{K}},t)-n_{-,\xi}(\vec{\mathrm{K}},t)\approx 1$; this decouples the SBEs in \equref{SBESecondForm}. Under this approximation, the interband current is~\cite{yue2022introduction,vampa2014theoretical,li2023high,chacon2020circular}
\begin{subequations}\label{eq:currentsKeldysh}
\begin{align}
\begin{split}\label{eq:Jinterkeldysh}
\bm{\mathrm{J}}^{(\xi)}_{\mathrm{inter}}(t)
&= -i\frac{d}{dt}\int d^2\vec{K}\,
\vec{d}^{(\xi)}_{+-}(\vec{K}+\vec{A}(t))
\int_{t_0}^{t}dt'\vec{E}(t')\cdot\left[\vec{d}^{(\xi)}_{+-}(\vec{K}+\vec{A}(t'))\right]^*\\
&\quad \times e^{-iS^{(\xi)}(\vec{K},t,t')-i(t-t')/T_2}
+\text{c.c.}
\end{split}\\
S^{(\xi)}(\vec{K},t,t')&= \int_{t'}^{t} dt''\Delta\epsilon(\vec{\vec{K}}+\vec{A}(t'')) + \vec{E}(t'')\cdot\Delta \vec{\mathrm{\mathcal{A}}}^{(\xi)}(\vec{K}+\vec{A}(t'')), \label{eq:S}
\end{align}\label{BothKeldyshEquations}
\end{subequations}
where $S(\vec{K},t,t')$ is the accumulated phase in time, also called the classical action. This solution is typically derived in the Houston basis and in the velocity gauge~\cite{vampa2014theoretical}, but this is a gauge invariant quantity as discussed in~\cite{yue2022introduction,li2019phase,chacon2020circular}. 

Here we examine \equref{eq:currentsKeldysh}, to qualitatively understand how the incident laser periodicity and the symmetry of the dispersion and matrix elements lead to the dominance of odd harmonic orders in this model. The HHG spectrum is proportional to the Fourier transform of the current, cf.~\equref{eq:HHG}. Basically, our argument will be that if \equref{eq:Jinterkeldysh} is effectively even in $\vec{A}(t)$, this would result in an HHG spectrum with peaks at odd frequencies, due to the extra factor of $\vec{E}(t)$ in \equref{eq:Jinterkeldysh}. 

We next discuss this more systematically. First, we consider how the instantaneous crystal momentum transforms in the moving frame, $\mathbf{k}(\mathbf{K},t)=\mathbf{K}+\mathbf{A}(t)$, under inversion of momentum $\vec{K}\rightarrow-\vec{K}$. For a monochromatic incident field, with frequency $\omega_0$ and period $\tau=2\pi/\omega_0$, the vector potential and electromagnetic field have approximately the following half-period symmetry (in the limit where $\tau$ is much shorter than the duration of the Gaussian envelope)
\begin{equation}
\mathbf{A}(t+\tau/2)\sim-\mathbf{A}(t),\qquad \mathbf{E}(t+\tau/2)\sim-\mathbf{E}(t).
\end{equation}
Including the half-period time shift in the moving frame, $\mathbf{K}+\mathbf{A}(t)$, gives
\begin{equation}
\mathbf{k}(\mathbf{K},t+\tau/2)\sim\mathbf{K}-\mathbf{A}(t)\sim-\mathbf{k}(-\mathbf{K},t).
\end{equation} 
Upon shifting $\vec{K} \rightarrow -\vec{K}$ in the momentum integration in  \equref{BothKeldyshEquations}, we can relate $\bm{\mathrm{J}}^{(\xi)}_{\mathrm{inter}}(t)$ to $\bm{\mathrm{J}}^{(\xi)}_{\mathrm{inter}}(t+\tau/2)$, where all functions are evaluated at $-\vec{k}$. As a consequence, by exploring the parity of each element in the right hand side \equref{eq:Jinterkeldysh} under momentum and half-period transformation in time, i.e. $\vec{k}\rightarrow-\vec{k}$ and $t\rightarrow t+\tau/2$, we can investigate the behavior of $\bm{\mathrm{J}}^{(\xi)}_{\mathrm{inter}}(t)$ under $t\rightarrow t+\tau/2$. We will discuss the parity of the following quantities: $\vec{d}^{(\xi)}_{+-}(\vec{k})$, $\mathbf{E}(t)$, $\Delta\epsilon(\vec{k})$ and $\Delta \vec{\mathrm{\mathcal{A}}}(\vec{k})$.

The parity of $\vec{d}^{(\xi)}_{+-}(\vec{k})$ is opposite to the parity of $n$, $\vec{d}^{(\xi)}_{+-}(-\vec{k})=-(-1)^{-n}\vec{d}^{(\xi)}_{+-}(\vec{k})$, therefore, an even (odd) $n$ gives an odd (even) $\vec{d}^{(\xi)}_{+-}(\vec{k})$. Nevertheless, in \equref{eq:Jinterkeldysh} the time-dependent current is proportional to $(\vec{d}^{(\xi)}_{+-}(\vec{k}))^2$, and this quantity has always even parity regardless of the parity of $n$, as a consequence, we can safely assume the parity of the harmonic orders will not depend on the parity of $n$.

Analyzing the parity of the terms in \equref{eq:S}, $\Delta\epsilon(\vec{k})$ is even [cf.~\equref{eq:Energy}] and $\Delta\vec{\mathrm{\mathcal{A}}}^{(\xi)}(\vec{k})$ is odd as the diagonal elements of $\vec{\mathrm{\mathcal{A}}}^{(\xi)}_p(\vec{k})$ are both odd under momentum inversion symmetry, following the relation $\vec{\mathrm{\mathcal{A}}}^{(\xi)}_p(\vec{k})=-\vec{\mathrm{\mathcal{A}}}^{(\xi)}_p(-\vec{k})$. Moreover, the  parity of $\vec{E}(t+\tau/2)$ is odd, transforming the parity of the product $\Delta\vec{\mathrm{\mathcal{A}}}^{(\xi)}(\vec{k})\vec{E}(t)$ into an even function. Resulting in an even function $S^{(\xi)}(\vec{k},t,t')$ under time and momentum transformations.

Taken together, $S^{(\xi)}(\vec{k},t,t')$ and $(\vec{d}_{+-}^{(\xi)}(\vec{k}))^2$ terms in \equref{eq:Jinterkeldysh} have even parity, and $\vec{E}(t)$ is an odd function, resulting into an odd function on the right hand side. Note that the time derivative does not affect this property since $g(t) = -g(t+\tau/2)$ implies that also $g'(t) = -g'(t+\tau/2)$.  
The fact that the current in \equref{eq:Jinterkeldysh} is odd under $t \rightarrow t + \tau/2$, will select the harmonic orders of type $(2m\pm1)\omega_0$, leading to a $\mathrm{FT}[\vec{J}_{\mathrm{inter}}^{(\xi)}(t)]$ to have only odd harmonic orders. This is consistent with a naive analysis of the current periodicity in time, i.e. if the current satisfies $\textbf{J}^{(\xi)}(t+\tau)=\textbf{J}^{(\xi)}(t)$ and $\textbf{J}^{(\xi)}(t+\tau/2)=-\textbf{J}^{(\xi)}(t)$, then it can be expanded as
\begin{equation}
\textbf{J}^{(\xi)}(t)=\sum_{m\in\mathbb Z} \textbf{J}_m e^{-im\omega_0 t} \quad\text{and} \quad \sum_m \textbf{J}_m (-1)^m e^{-im\omega_0 t}=-\sum_m \textbf{J}_m e^{-im\omega_0 t} \ \  \Rightarrow \ \ [(-1)^m+1]\textbf{J}_m=0,
\label{FourierExpansion}
\end{equation}
resulting in non-zero Fourier coefficients only for odd $m$. Nevertheless, in a finite-duration Gaussian pulse, containing several cycles of the main carrier frequency $\omega_0$, this symmetry is only approximate, which is consistent with the strong suppression, rather than the exact absence, of even harmonics in our simulations.

\subsection{A qualitative argument for the dominance of $2n \pm 1$}
We now search for an intuitive explanation of the scaling of the interband HOs with $n$. We will focus on valley $\xi=+$ and RCP, for simplicity. The interband dipole matrix element vector is
\begin{equation}
\mathbf{d}_{+-}(\mathbf{k})=\big(d^x_{+-}(\mathbf{k}),\, d^y_{+-}(\mathbf{k})\big),
\end{equation}
which we can write in its circular components
\begin{equation}
d_{+-}^{\pm}(\mathbf{k}) := d^x_{+-}(\mathbf{k}) \pm i\, d^y_{+-}(\mathbf{k}).
\end{equation}
Similarly, for the electric field we define the circular polarization basis with $E_\pm$, resulting in the expression
\begin{equation}
E_\pm(t):=E_x(t)\pm\,iE_y(t)\propto e^{\mp i \omega_0(t - t_0)} .
\end{equation}
From Eq.~(\ref{d+-}), both components of the interband dipole contain a common factor $(k_x+ik_y)^n$ multiplied by a term linear in $k_x$ and $k_y$. Passing to circular components simplifies this structure further: the combination $d^x_{+-}+i d^y_{+-}$ produces one additional factor $(k_x+ik_y)$, whereas $d^x_{+-}-i d^y_{+-}$ produces one factor $(k_x-ik_y)$, which can be rewritten as $k^2/(k_x+ik_y)$. Therefore, the two circular components can be written as
\begin{equation}
d_{+-}^{+}(\mathbf{k})=\mathcal{D}_{+}(k)\,(k_x+ik_y)^{n+1},
\qquad
d_{+-}^{-}(\mathbf{k})=\mathcal{D}_{-}(k)\,(k_x+ik_y)^{n-1},
\end{equation}
where $\mathcal{D}_{\pm}(k)$ are radial prefactors. Writing
\begin{equation}
k_x(t)+ik_y(t)=k(t)e^{i\varphi_{\mathbf{k}}(t)},
\end{equation}
where $\varphi_{\mathbf{k}}(t)$ is the polar angle of the instantaneous momentum $\mathbf{k}(t)=\mathbf{K}+\mathbf{A}(t)$, one obtains
\begin{equation}
d_{+-}^{+}(\mathbf{k}(t))\propto e^{i(n+1)\varphi_{\mathbf{k}}(t)},
\qquad
d_{+-}^{-}(\mathbf{k}(t))\propto e^{i(n-1)\varphi_{\mathbf{k}}(t)}.
\end{equation}
Thus, the two circular components of the interband dipole carry angular windings $n+1$ and $n-1$, respectively.

The interband coherence $\pi_\xi(\mathbf{K},t)$ is driven by the source term
\begin{equation}
\mathbf{E}(t)\cdot \mathbf{d}_{+-}(\mathbf{k}(t))
=\frac{1}{2}\Big(E_+(t)\, d_{+-}^{+}(\mathbf{k}(t)) + E_-(t)\, d_{+-}^{-}(\mathbf{k}(t))\Big).
\end{equation}
For a nearly monochromatic circular pulse, the field-driven part $\mathbf{A}(t)$ rotates with frequency $\omega_0$ during the central cycles of the pulse. We assume for those trajectories and times dominating the interband response that the angle $\varphi_{\mathbf{k}}(t)$ therefore evolves approximately linearly in time,
$
\varphi_{\mathbf{k}}(t)\approx \omega_0 t+\varphi_0.
$
The circular components of the field carry one unit of angular phase,
$
E_\pm(t)\propto e^{\mp i\omega_0 t}\approx e^{\mp i\varphi_{\mathbf{k}}(t)}$ such that the two contributions to the source term behave as
\begin{equation}
\mathbf{E}(t)\cdot \mathbf{d}_{+-}(\mathbf{k}(t)) = \frac{1}{2}[E_+(t)\cdot d^+_{+-}(\mathbf{k}(t)) + E_-(t)\cdot d^-_{+-}(\mathbf{k}(t))]
 \propto e^{in\varphi_{\mathbf{k}}(t)}.
\end{equation}
Hence, irrespective of the helicity channel in the driving term, the interband source carries an $n$-fold angular winding. In this semiclassical central-cycle picture, the interband coherence therefore acquires a dominant component with the same winding,
$
\pi_\xi(\mathbf{K},t)\propto e^{in\varphi_{\mathbf{k}}(t)}.
$
When inserted back into the interband current, \equref{eq:Jinter}, the additional dipole factor has a phase with winding $n\pm1$. Consequently, the emitted interband response contains only the neighboring odd orders $2n\pm1$.

\section{Additional numerical results}
\label{sec:AppHHG}

\begin{figure}[!h]
    \centering
    \includegraphics[width=0.35\columnwidth]{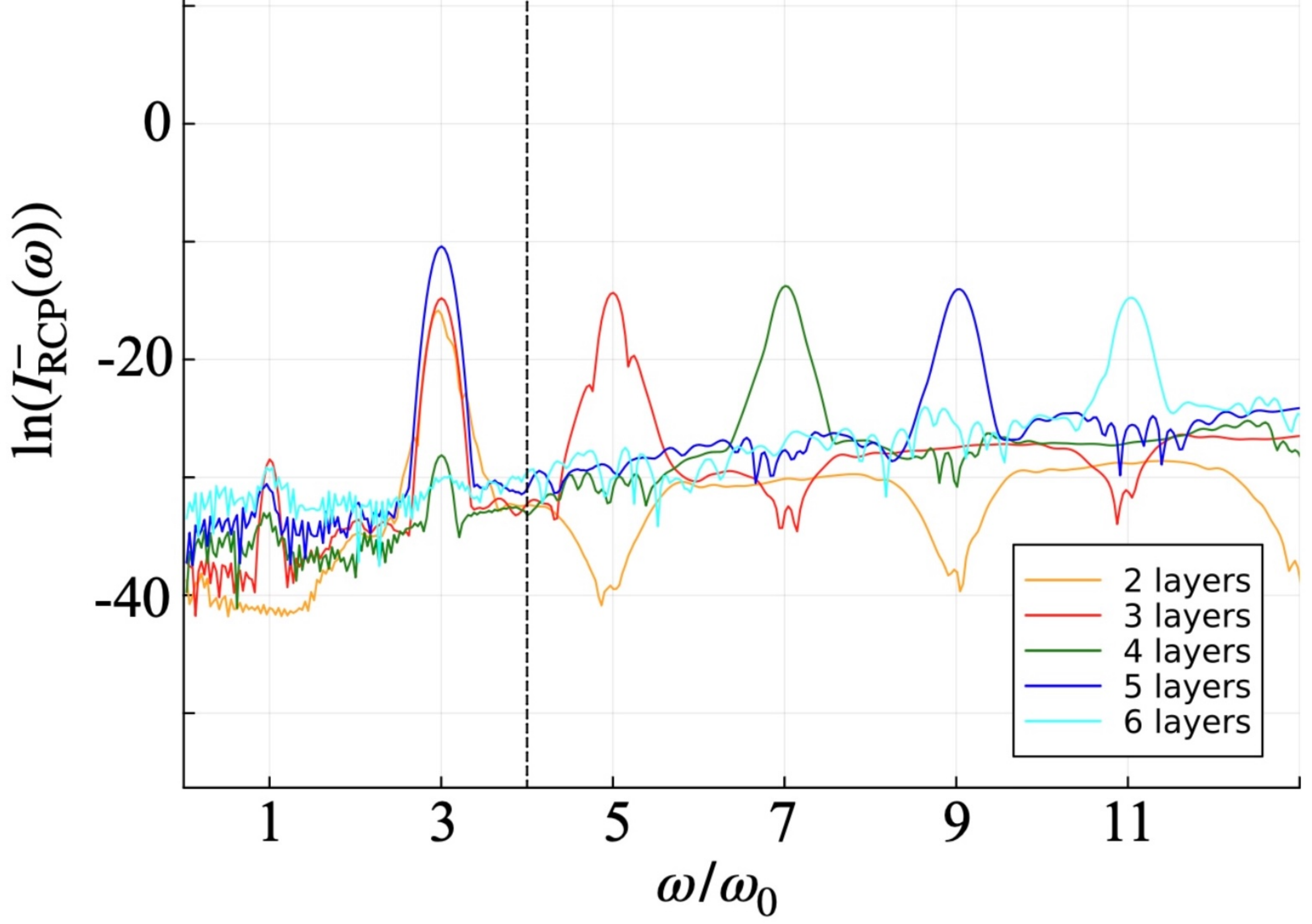}
    \caption{$I^-_{\mathrm{RCP}}(\omega)$ as function of $\omega/\omega_0$ for $n=2-6$.} 
    \label{fig:ImRCP}
\end{figure}

In \figref{fig:ImRCP} we show the HHG of $I^-_{\mathrm{RCP}}(\omega)$. Here the dominant HOs have dominant peaks at $2n-1$. Similarly, the same HOs are dominant for $I^+_{\mathrm{LCP}}(\omega)$. Nevertheless, these do not contribute to $\mathrm{CD}_l$ in \equref{eq:CD}.

\begin{figure}[!h]
    \centering
    \includegraphics[width=0.35\columnwidth]{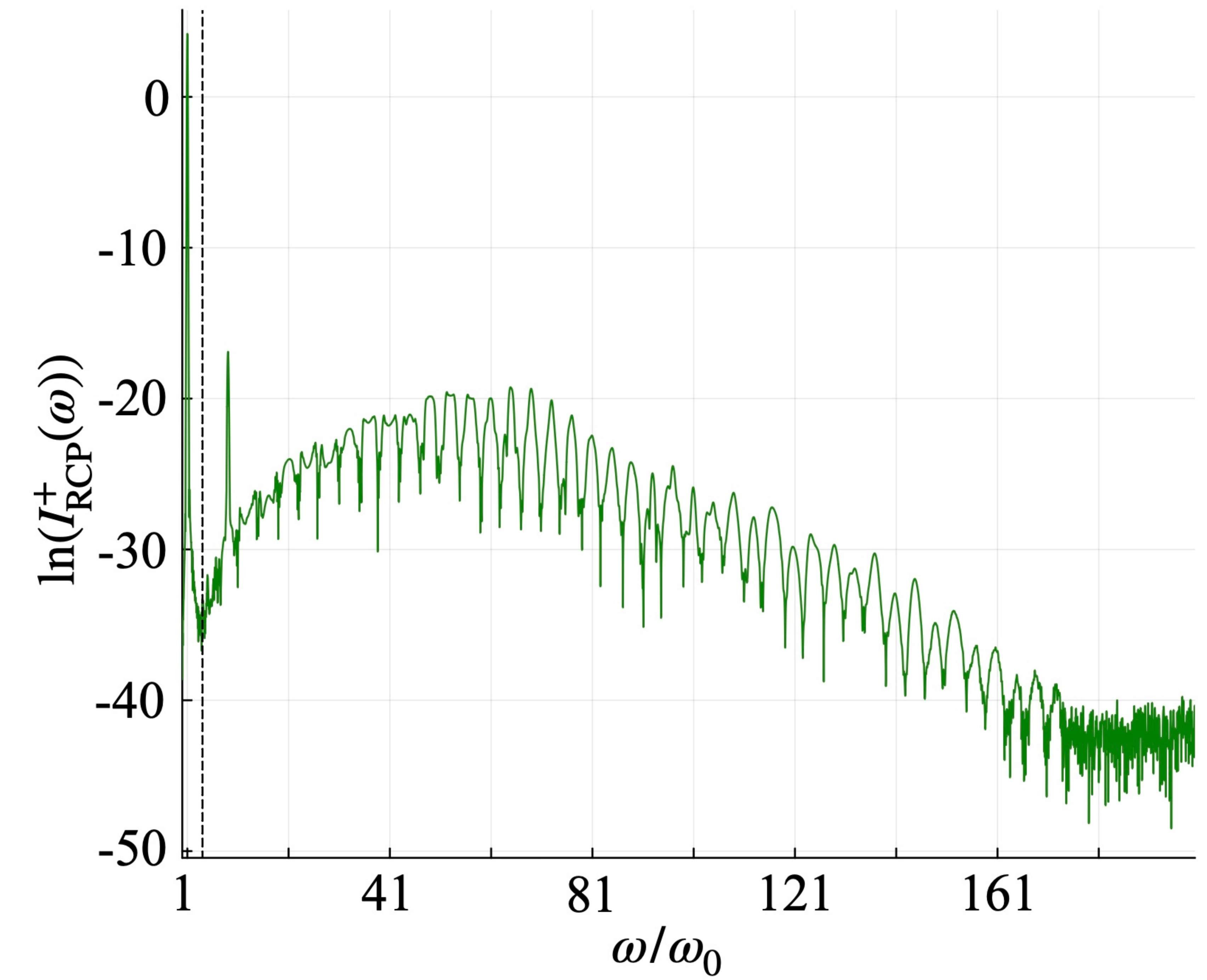}
    \caption{HHG spectra for $n=4$ under RCP.} 
    \label{fig:HHGAll}
\end{figure}

In \figref{fig:HHGAll} we show the full HHG spectra of $I^+_{\mathrm{RCP}}(\omega)$, with the behavior of higher HOs of the HHG spectra, including plateau and cutoff regions, as is typical in HHG. 
\newpage

\section{Additional results on the relative Current distribution}
\label{sec:AppCurCircular}
\begin{figure}[!h]
    \centering
    \includegraphics[width=0.5\columnwidth]{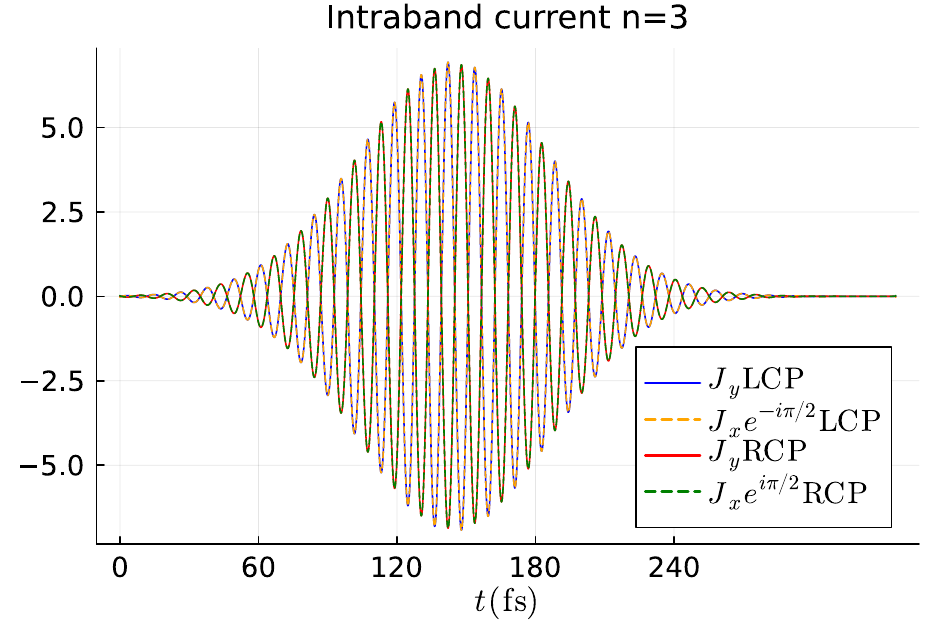}
    \caption{Intraband current for $n=3$ for $J_x$, $J_y(t)$ components under RCP and LCP. From the current $J_x(t)$ we extract a function $A(t) e^{i(\theta(t))}$, to which we add a phase shift $e^{- i\epsilon\pi/2}$. Shown in the figure $\mathrm{Re}[A(t) e^{i(\theta(t)\pm\pi/2)}]$, for a single valley $\xi=+$. The phase shift as a function of the ellipticity $\epsilon$ is $e^{- i\epsilon\pi/2}$.} 
    \label{fig:CurIntra3}
\end{figure}

\begin{figure}[!h]
    \centering
    \includegraphics[width=0.6\columnwidth]{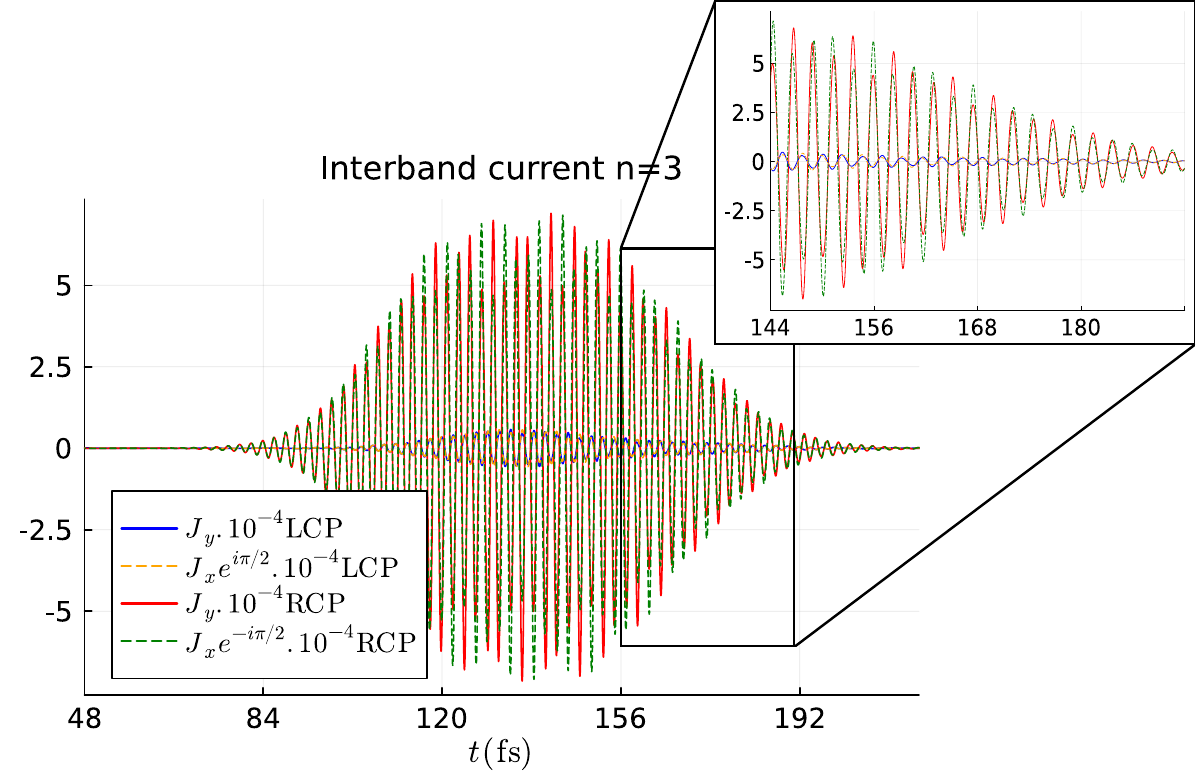}
    \caption{Interband current for $n=3$ for $J_x(t)$, $J_y(t)$ components under RCP and LCP. $J_x(t)$ is modified under a $e^{i\epsilon\pi/2}$ phase shift. The figure shows the real part of $A(t) e^{i(\theta(t)\pm\pi/2)}$, for a single valley  in valley $\xi=+$.The phase shift as a function of the ellipticity $\epsilon$ is $e^{i\epsilon\pi/2}$.} 
    \label{fig:CurInter3}
\end{figure}

Here we compare the relative phase between $\hat{x}$ and $\hat{y}$ components of currents under right (left) circular polarization showing how a they have a similar oscillatory pattern, up to a relative $\pi/2$ phase shift. Figs.~\ref{fig:CurIntra3},\ref{fig:CurInter3} show the results for interband and intraband for $n=3$, similar phase relations holds for $n=2-6$. 

From the $\hat{x}$-component current distribution $J_x(t)$ we extract a function with pattern $A(t) e^{i\theta(t)}$, to which we add a $\pm\pi/2$ phase shift, transforming to $A(t) e^{i(\theta(t)\pm\pi/2)}$, here we plot the real part as a function of time in Figs.~\ref{fig:CurIntra3},\ref{fig:CurInter3}. The results for the intraband current are shown in \figref{fig:CurIntra3}, where we include a relative phase in the $\hat{x}$ component depending on the ellipticity $\epsilon$ as $e^{- i\epsilon\pi/2}$. The phase shifted $\hat{x}$ component is in phase with the $\hat{y}$ component. This is the same phase relation between the components of the input laser beam. The results for the interband current are shown in \figref{fig:CurInter3}, here the relative phase included in the $\hat{x}$ component is $e^{i\epsilon\pi/2}$. The phase relation between the $\hat{x}$ and $\hat{y}$ components of interband current is opposite to the phase relation of the intraband current and the input laser components.

\newpage

\begin{figure}[h!]
    \centering
    \includegraphics[width=0.55\columnwidth]{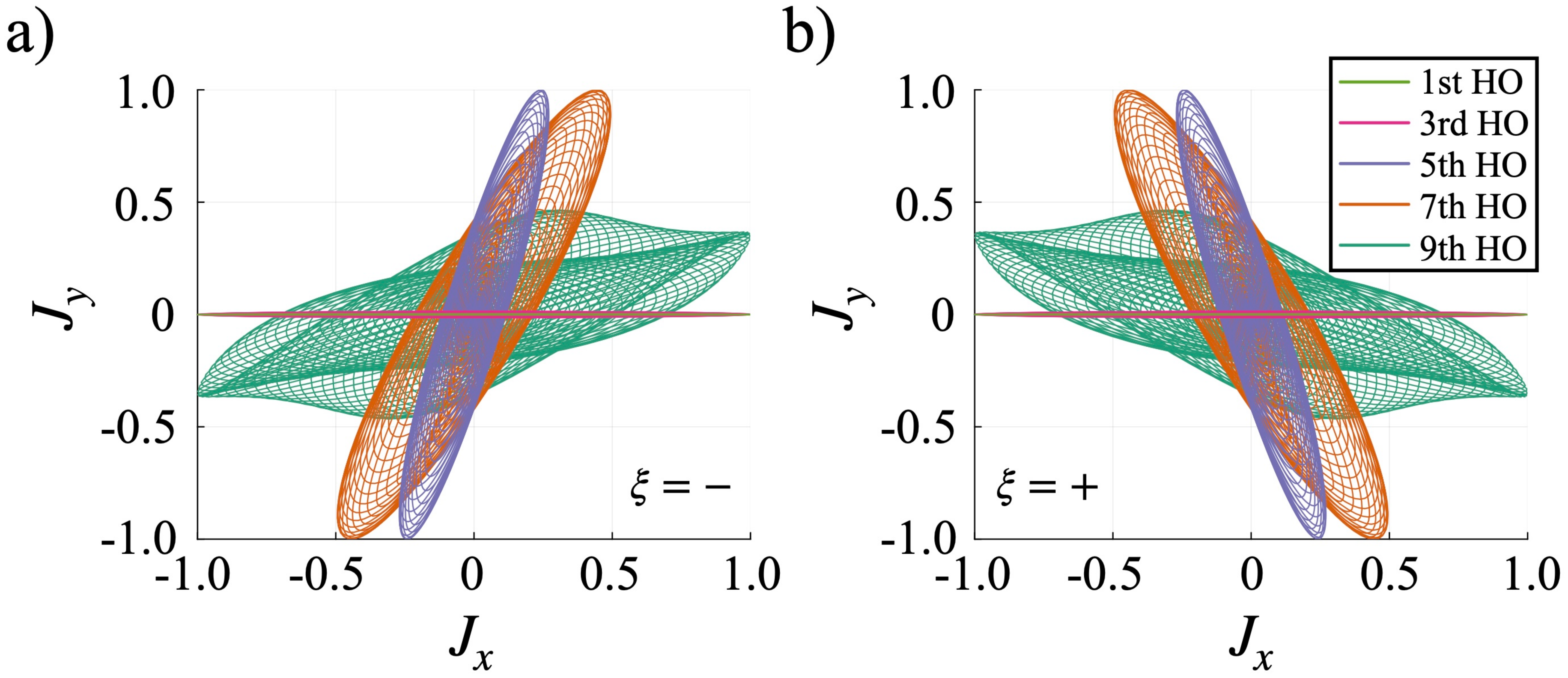}
    \caption{$J_x$,$J_y$ relative currents at each harmonic order for $n=4$, considering each valley $\xi=\pm$ individually. a) $\xi=-$valley. b) $\xi=+$ valley. }
    \label{fig:2VPpolHO}
\end{figure}

In \figref{fig:2VPpolHO} we present similar results as in \figref{fig:2VP}, but without the $2$ valley approximation. Each valley has the same gap but opposite chirality and are independently interacting with the incident field in the $\hat{x}$ direction. We analyze the results at each harmonic order. Showing how each valley chirality affects the polarization. Above the bandgap the rotation is in agreement with the mean field $2$ valley approximation orientation, valley $\xi=-$, left, ($\xi=+$, right), the polarization is slightly rotated to the left (right) of the incident polarization. In more details, this is calculating computing a Fourier transform the time dependent current $\tilde{J}(\omega)=\mathrm{FT}[J(t)]$, a Gaussian mask around each HO, followed by a Fourier transform back to time domain $J_l(t)=\mathrm{FT}[e^{-(\omega-\omega_l)^2/\chi^2}\tilde{J}(\omega)]$, where $\chi$ is the width of the HO. 
\end{appendix}

\end{document}